\documentclass[journal,10pt]{IEEEtran}
\usepackage{makecell}
\usepackage{tabularx}
\usepackage{graphics}
\usepackage{graphicx}
\usepackage{subfig}
\usepackage{amsmath,amsfonts,amssymb,varioref,mathrsfs,verbatim}
\usepackage{latexsym}
\usepackage{epsfig}
\usepackage{array}
\usepackage[english]{babel}
\usepackage{color}
\usepackage{babel}

\newtheorem{lemma}{Lemma}
\newtheorem{remark}{Remark}

\renewcommand{\figurename}{Fig.}
\addto\captionsenglish{\renewcommand{\figurename}{Fig.}}
\makeatletter
\let\l@ENGLISH\l@english
\makeatother
\begin{document}
%\title{On the Privacy and Security of Millimeter Wave Vehicular Communication Networks}
\title{Enhancing Secrecy with Multi-Antenna Transmission in Millimeter Wave Vehicular Communication Systems}
\author{Mohammed E. Eltayeb, Junil Choi, Tareq Y. Al-Naffouri, and Robert~W.~Heath,~Jr. 

\thanks{Mohammed E. Eltayeb, Junil Choi, and Robert W. Heath, Jr. are with The University of Texas at Austin (Email: meltayeb,rheath@utexas.edu). Tareq Y. Al-Naffouri  is with King Abdullah University of Science and Technology (Email: tareq.alnaffouri@kaust.edu.sa).  This research was partially supported by the U.S. Department of Transportation through the Data-Supported Transportation Operations and Planning (D-STOP) Tier 1 University Transportation Center and by the Texas Department of Transportation under Project 0-6877 entitled ��Communications and Radar-Supported Transportation Operations and Planning (CAR-STOP).}
}
\maketitle

%\markboth{IEEE Transactions on Vehicular Technology,~Vol.~XX, No.~XX, XXX~2016}
\IEEEpeerreviewmaketitle

\begin{abstract}

Millimeter wave (mmWave) vehicular communication systems will provide an abundance of bandwidth for the exchange of raw sensor data and support driver-assisted and safety-related functionalities. Lack of secure communication links, however, may lead to  abuses and attacks that jeopardize the efficiency of transportation systems and the physical safety of drivers. In this paper, we propose two physical layer (PHY)  security techniques for vehicular mmWave communication systems. The first technique uses multiple antennas with a single RF chain to transmit information symbols to a target receiver and noise-like signals in non-receiver directions. The second technique uses multiple antennas with a few RF chains to transmit information symbols to a target receiver and opportunistically inject artificial noise in controlled directions, thereby reducing interference in vehicular environments.   Theoretical and numerical results show that the proposed techniques provide higher secrecy rate when compared to traditional PHY security techniques that require  digital or more complex antenna architectures.

%By deriving closed form expressions for the secrecy rate, we show that the proposed techniques provide higher secrecy rate when compared to traditional PHY security techniques that require  digital or more complex antenna architectures.
%show that the proposed techniques provide higher secrecy rate when compared to traditional PHY security techniques that require  digital or more complex antenna architectures.
%The large dimensional antenna arrays available in mmWave systems can be used to develop low-complexity and effective physical layer (PHY) security techniques.
%The high power consumption of the mixed signal components at the mmWave band, however, makes the implementation of traditional PHY security techniques in mmWave systems challenging.
\end{abstract}

%%%%%%%%%%%%%%%%%%%%%%%%%%%%%%%%%%%%%%%%%%%%%%%%%%%%%%%%%%%%%%%%%%%%%%%%%%%%%%%%%%%%%%%%%%%%%%%%%%%%%%%%%%%%%
\section{Introduction} \label{sec:Intro}
%%%%%%%%%%%%%%%%%%%%%%%%%%%%%%%%%%%%%%%%%%%%%%%%%%%%%%%%%%%%%%%%%%%%%%%%%%%%%%%%%%%%%%%%%%%%%%%%%%%%%%%%%%%%%%

Millimeter wave (mmWave) communication is considered one of the key enabling technology to provide high-speed wideband vehicular communication and enable a variety of applications for safety, traffic efficiency, driver assistance, and infotainment \cite{r1}, \cite{r0}. Like any communication system, vehicular communication systems are vulnerable to various security threats that could jeopardize the efficiency of transportation systems and the physical safety of vehicles and drivers.  Typical threats include  spoofing \cite{r4}, message falsification attacks \cite{r2},  message reply attacks \cite{r4c}, eavesdropping \cite{r5},  and many others \cite{r4}, \cite{r6}. 

 A number of encryption protocols based on digital signature techniques have been proposed in the literature  to preserve the privacy and security of lower frequency vehicular communication systems \cite{r7}-\cite{r9}.  The  shift from lower frequency to higher frequency mmWave systems, however, introduces new challenges that cannot be fully handled by traditional cryptographic means. For instance, in lower frequency systems, public cryptographic keys can be simply broadcasted over the wireless channel, however, in mmWave systems, the directionality of the mmWave channel dictates dedicated public key transmission to all network nodes. This imposes a great deal of overhead on the system, and might not be an option for applications that require strict latencies for message delivery or are time-sensitive.  Moreover, these cryptographic techniques generally require an infrastructure for key distribution and management which might not be readily available \cite{r2}, \cite{r6}. With the emergence of new time-sensitive  safety applications and the increasing size  (more than two billion by 2020 \cite{bc}) and heterogeneity of the decentralized vehicular network, the implementation of traditional cryptographic techniques becomes complex and challenging.

To mitigate these challenges,  \textit{keyless} physical layer (PHY) security techniques \cite{an1}-\cite{dm5}, which do not directly rely on upper-layer data encryption or secret keys, can be employed to secure vehicular communication links. These techniques use multiple antennas at the transmitter to generate the desired information symbol at the receiver and  \textit{artificial noise} at potential eavesdroppers. The precoder design is based on the use of digital beamforming  \cite{an1}-\cite{anr1}, distributed arrays (relays) \cite{dm403}, \cite{anr1}, and/or switched arrays \cite{s1}-\cite{dm5}. In the digital beamforming techniques  \cite{an1}-\cite{anr1}, antenna weights are designed so that artificial noise can be uniformly injected into the null space of the receiver's channel to jam potential eavesdroppers. In the distributed array techniques \cite{dm403}, \cite{anr1}, multiple relays are used to jointly  aid communication from the transmitter to the receiver and result in artificial noise at potential eavesdroppers. In the switched array techniques \cite{s1}-\cite{dm5},  a set of antennas, co-phased towards the intended receiver, are randomly associated with every transmission symbol. This results in coherent transmission towards the receiver and induces artificial noise in all other directions.

Despite their effectiveness, the techniques proposed in \cite{an1}-\cite{anr1} are not suitable for mmWave systems. The high power consumption of the mixed signal components at the mmWave band makes it difficult to allocate an RF chain for each antenna \cite{pi2011}. This restricts the feasible set of antenna weights that can be applied and makes the precoding design challenging. While the switched array techniques \cite{s1}-\cite{dm5}  can be implemented in mmWave systems, it was recently shown that the antenna sparsity, as a result of switching, can be exploited to launch plaintext and other attacks as highlighted in \cite{attack1}, \cite{attack2}. In addition to the limitations, the road geometry restricts the locations of the potential eavesdroppers in vehicular environments. This prior information was not exploited in previous techniques and can be used to opportunistically inject artificial noise in a controlled direction along the lane of travel.

In this paper, we propose two PHY security techniques for vehicular mmWave communication systems that (i) do not require the exchange of secret keys among vehicles as done in \cite{r7}-\cite{r9}, and (ii) do not require fully digital antenna architectures as done in \cite{an1}-\cite{anr1}. In the first technique, a phased array with a single RF chain is used to design analog beamformers. Specifically, a random subset of antennas is used to beamform information symbols to the receiver, while the remaining antennas are used to randomize the far field radiation pattern at non-receiver directions. This results in coherent transmission to the receiver and a noise-like signal that jams potential eavesdroppers with sensitive receivers.  The proposed technique is different from the switched array technique proposed in \cite{dm5}, which  also distorts the far field radiation pattern at non-receiver directions, since: (i) the switched array technique \cite{dm5} uses antenna switches to select a few antennas for beamforming and the remaining antennas are idle, and (ii) the idle antennas in \cite{dm5} create a sparse array which could be exploited by adversaries to estimate and precancel the sidelobe distortion \cite{attack1}, \cite{attack2}. The proposed technique reduces the antenna complexity since it does not require antenna switches, and the design criteria and analytical approaches undertaken in this work are different from \cite{dm5}. Moreover,  there are no idle antennas in the proposed technique, thus making it difficult to breach.

%The proposed technique  uses all antennas to distort the sidelobes, therefore making it difficult to breach.  Moreover, 

In the second technique, a phased array with a few RF chains is used to design hybrid analog/digital transmission precoders. The hybrid precoders are designed to beamform information symbols towards the receiver and simultaneously inject artificial noise in directions of interest.  Prior road network information, e.g. boundaries, lane width, junctions, etc., is used  to opportunistically inject controlled artificial noise in threat directions instead of spreading the artificial noise in all non-receiver directions as done in \cite{an1}-\cite{anr1}. This has two immediate advantages: (i) it allows the transmitter to exploit its antenna array gain when generating artificial noise, and (ii) it reduces interference to friendly or non-threat regions. The road network information can be obtained from an onboard camera, radar or a geographical information system.

The proposed security technique using the hybrid design is different from the techniques proposed in \cite{an1}-\cite{anr1}, which also inject artificial noise in non-receiver directions, since these techniques: (i) require a fully digital antenna architecture, (ii) do not have the ability to control the directions of the injected noise, and (iii) do not exploit the array gain in the generation of the artificial noise. Although hybrid designs were undertaken in \cite{mol}-\cite{sw}, these designs primarily targeted cellular systems.  Moreover, the design in \cite{sw} relies on antenna subset selection. This reduces the array gain since only a few antennas are active when compared to the designs in \cite{ahmed1}, \cite{omar}. The hybrid design introduced in this paper requires fewer RF chains when compared to the hybrid designs undertaken in \cite{mol}-\cite{sw}, to achieve  performance close to that obtained by digital architectures.

The remainder of this paper is organized as follows. In Section \ref{sec:Model}, we introduce the system and channel model. In Sections \ref{sec:ST} and \ref{sec:ST2}, we present the proposed PHY security techniques. In Section \ref{sec:ana} we analyze the performance of the proposed techniques and provide numerical results and discussions in Section \ref{sec:num}. We conclude our work in Section \ref{sec:con}.

		\begin{figure}[t]
			\begin{center}
			\vspace{2mm}
	\includegraphics[width=2.8in]{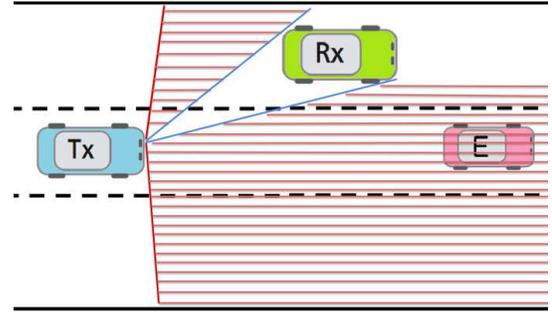}
		\caption{Vehicle-to-vehicle communication with a possible eavesdropper. The transmitting vehicle (Tx)  is communicating with the target vehicle (Rx) while the eavesdropper (E) tries to intercept the transmitted data. Tx jams regions with horizontal lines.}
					\label{fig:mod4}
			\end{center}
		\end{figure}

%%%%%%%%%%%%%%%%%%%%%%%%%%%%%%%%%%%%%%%%%%%%%%%%%%%%%%%%%%%%%%%%%%%%%%%%%%%%%%%%%%%%%%%%%%%%%%%%%%%%%%%%%%%%%
\section{System and Channel Model} \label{sec:Model}
\begin{figure}[t]
		\begin{center}
\includegraphics[width=2.8in]{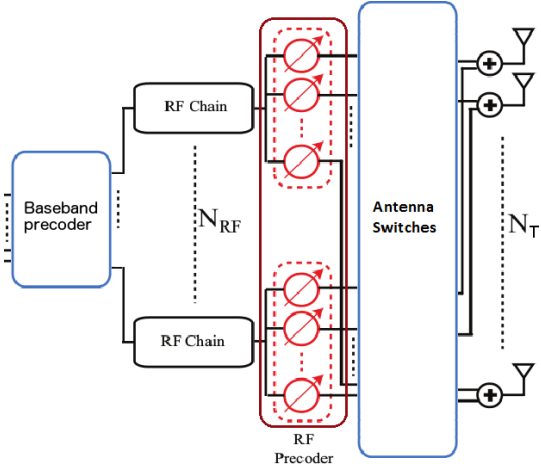}
				\caption{Proposed hybrid architecture where the number of antennas $N_\text{T}$ are much larger than the number of RF chains.  In the case of a single RF chain, the antenna architecture simplifies to the conventional analog architecture.}
			\label{fig:model}
		\end{center}
	\end{figure}

%%%%%%%%%%%%%%%%%%%%%%%%%%%%%%%%%%%%%%%%%%%%%%%%%%%%%%%%%%%%%%%%%%%%%%%%%%%%%%%%%%%%%%%%%%%%%%%%%%%%%%%%%%%%%
We consider a mmWave transmitting vehicle (transmitter) and a receiving vehicle (receiver) in the presence of a single or multiple eavesdroppers with sensitive receivers as shown in Fig. \ref{fig:mod4}. The transmitter is equipped with $N_{\text{T}}$ transmit antennas and $N_\text{RF}$ RF chains as shown in Fig. \ref{fig:model}.  In the case of a single RF chain, the  architecture in Fig. \ref{fig:model} simplifies to the conventional analog architecture. We adopt a uniform linear array (ULA) with isotropic antennas along the x-axis with the array centered at the origin; nonetheless, the proposed technique can be adapted to other antenna structures. This can be done by, for example, varying the complex antenna weights to form constructive and destructive interference as described in the Section \ref{sec:ST} or by using a digital or hybrid architecture as shown in Fig. \ref{fig:model}. Since the array is located along the x-y plane, the receiver's location is specified by the azimuth angle of arrival/departure (AoA/AoD) \cite{trees}. We consider a single wiretap channel where the transmitter is assumed to know the angular location of the target receiver via beam training, but not of the potential eavesdroppers.

%&=& \sqrt{P N_\text{R}\alpha}\mathbf{h}^*(\theta)\mathbf{f}s(k) + z(k),

The transmit data symbol $s[k]\in \mathbb{C}$, where $\mathbb{E}[|s(k)|^2]=1$ and $k$ is the symbol index,  is multiplied by a unit norm transmit beamforming/precoding vector $\mathbf{f} = [f_1 \quad f_2 \quad ... \quad f_{N_\text{T}}]^{\mathrm{T}}\in \mathbb{C}^{N_{\text{T}}} $ with $f_n$ denoting the complex weight on transmit antenna $n$. At the receiver,  the received signals on all antennas are combined with a receive combining unit norm vector $w = [w_1 \quad w_2 \quad...\quad w_{N_\text{R}}]^{\mathrm{T}}\in \mathbb{C}^{N_{\text{R}}}$, where $N_\text{R}$ is the number of antennas at the receiver. We assume a narrow band line-of-sight (LOS) channel with perfect synchronization between the transmitter and the receiver.

Due to the dominant reflected path from the road surface, a two-ray model is usually adopted in the literature to model LOS vehicle-to-vehicle communication \cite{cv1}-\cite{cv4}. Based on this model, the received signal can be written as  \cite{cv1}-\cite{cv4}
\begin{eqnarray}\label{c1mimo}
%y(k,\theta,\phi) = \sqrt{P\alpha}\mathbf{w}^*\mathbf{H}(\theta,\phi)\mathbf{f}s(k) + z(k),
y(k,\theta) = \sqrt{P\alpha}\mathbf{h}^*(\theta)\mathbf{f}s(k) + z(k),
\end{eqnarray}
where $k$ is the $k$th transmitted symbol, $\theta$ is AoD from the transmitter to the receiver,  $\alpha$ is the path-loss, $P$ is the transmission power, $\mathbf{h}^*(\theta)=\mathbf{w}^*\mathbf{H}(\theta,\phi)$ is the effective channel after combining,  and  $z(k) \sim  \mathcal {CN}(0,\sigma^2)$ is the additive noise.  The channel $\mathbf{H}(\theta,\phi)$ is given by
\begin{eqnarray}
\mathbf{H}(\theta,\phi) = g \mathbf{a}_\text{r}(\phi)\mathbf{a}^*_\text{}(\theta),
\end{eqnarray}
where the vectors $ \mathbf{a}_\text{r}(\phi)$ and $\mathbf{a}_\text{}(\theta)$ represent the receive and transmit array response vectors,  $\phi$ is the receiver's AoA, and the random variable $g$ captures the small scale fading due to multi-path and/or doppler spread. For sub-6 GHz frequencies, the distribution of $g$ is shown to be Gaussian in \cite{cv1}, however, to the best of the author's knowledge, the distribution of $g$ is unknown for 60 GHz frequencies. Nonetheless, it should be noted that beamforming at both the transmitter and receiver leads to a smaller Doppler spread and larger coherence time \cite{vuthaj}, \cite{d0}. Additionally, the angular variation is typically an order of magnitude slower than the conventional coherence time \cite{vuthaj}, therefore the overhead to align both transmit and receive beams is not substantial. Setting $\mathbf{w}=\frac{ \mathbf{a}_\text{r}(\phi)}{\sqrt{N_\text{R}}}$, the channel $\mathbf{h}^*(\theta)$ becomes $\mathbf{h}^*(\theta)= \sqrt{N_\text{R}}g\mathbf{a}^*_\text{}(\theta)$. For a ULA with $d \leq \frac{\lambda}{2}$ antenna spacing, the array response vectors  $ \mathbf{a}_\text{}(\theta)$ and $\mathbf{a}_\text{r}(\phi)$ are $\nonumber \mathbf{a}_\mathrm{}(\theta) = [e^{j( \frac{N_\text{T}-1}{2}) \frac{2\pi d}{\lambda} \cos (\theta)}, \quad e^{j( \frac{N_\text{T}-1}{2}-1) \frac{2\pi d}{\lambda} \cos (\theta)}, ...,\\ \quad e^{-j( \frac{N_\text{T}-1}{2}) \frac{2\pi d}{\lambda} \cos (\theta)}   ]^{\mathrm{T}}$, and $\mathbf{a}_\mathrm{r}(\phi) = [e^{j( \frac{N_\text{R}-1}{2}) \frac{2\pi d}{\lambda} \cos (\phi)}, \\ \quad e^{j( \frac{N_\text{R}-1}{2}-1) \frac{2\pi d}{\lambda} \cos (\phi)},..., \quad e^{-j( \frac{N_\text{R}-1}{2}) \frac{2\pi d}{\lambda} \cos (\phi)}   ]^{\mathrm{T}}$ \cite{trees}.

%%%%%%%%%%%%%%%%%%%%%%%%%%%%%%%%%%%%%%%%%%%%%%%%%%%%%%%%%%%%%%%%%%%%%%%%%%%%%%%%%%%%%%%%%%%%%%%%%%%%%%%%%%%%%
\section{Analog Beamforming with Artificial Jamming} \label{sec:ST}
%%%%%%%%%%%%%%%%%%%%%%%%%%%%%%%%%%%%%%%%%%%%%%%%%%%%%%%%%%%%%%%%%%%%%%%%%%%%%%%%%%%%%%%%%%%%%%%%%%%%%%%%%%%%%
In this section, we introduce an analog PHY security technique that randomizes the information symbols at potential eavesdroppers without the need for a fully digital array or antenna switches. We adopt the antenna architecture shown in Fig. \ref{fig:model} with a single RF chain. The antenna switches in Fig. \ref{fig:model} are  not required for the analog beamforming design. Instead of using all antennas for beamforming,  a random subset of antennas are set for coherent beamforming, while the remaining antennas are set to destructively combine at the receiver. The indices of these antennas are randomized in every symbol transmission. This randomizes the beam pattern sidelobes. Although the target receiver would observe gain reduction, malicious eavesdroppers will observe  non-resolvable noise-like signals. 
  
Let $\mathcal{I}_M(k)$ be a random subset of $M$ antennas, where $M$ is selected such that $N_\text{T}-M$ is an even number, used to transmit the $k$th symbol, $\mathcal{I}_L(k)$ be a subset that contains the indices of the remaining antennas, $\mathcal{E}_L(k)$ be a subset that contains the even entries of   $\mathcal{I}_L(k)$, and $\mathcal{O}_L(k)$ be a subset that contains the odd entries of $\mathcal{I}_L(k)$.  The $n$th entry of the analog beamforming vector $\mathbf{f}(k)$ is set as $f_n(k)=\frac{1}{\sqrt{N_\text{T}}}e^{ j \Upsilon_{n}(k)}$ where 
  \begin{equation}\label{efbp}
\Upsilon_{n}(k)  = \left\{
               \begin{array}{ll}
              { \left(\frac{N_\text{T}-1}{2}-n \right)2\pi \frac{d}{\lambda} \cos(\theta_{\text{R}})}, & \hbox{  $n\in \mathcal{I}_M(k)$  } \\
                 { \left(\frac{N_\text{T}-1}{2}-n \right)2\pi \frac{d}{\lambda} \cos(\theta_{\text{R}})}, & \hbox{  $n\in \mathcal{E}_L(k)$  }  \\
                 { \left(\frac{N_\text{T}-1}{2}-n \right)2\pi \frac{d}{\lambda} \cos(\theta_{\text{R}})}+\pi, & \hbox{  $n\in \mathcal{O}_L(k)$  } \\
               \end{array}
               \right.
\end{equation} 
where $\theta_{\text{R}}$ is the transmit direction towards the target receiver.  Note for $n~\in~\mathcal{O}_L(k),$
\begin{eqnarray}
\nonumber f_n(k)=\frac{1}{\sqrt{N_\text{T}}}e^{j({ \left(\frac{N_\text{T}-1}{2}-n \right)2\pi \frac{d}{\lambda} \cos(\theta_{\text{R}})}+\pi)} \\  \label{efbpx} = -\frac{1}{\sqrt{N_\text{T}}} e^{j({ \left(\frac{N_\text{T}-1}{2}-n \right)2\pi \frac{d}{\lambda} \cos(\theta_{\text{R}})})}.
\end{eqnarray}   
Using (\ref{c1mimo})-(\ref{efbpx}), the received signal  at an arbitrary  receiver becomes
\begin{eqnarray}\label{arnl}
\begin{aligned} 
%& y(k,\theta,\phi) = \sqrt{P\alpha}\mathbf{w}^*\mathbf{H}(\theta,\phi)\mathbf{f}(k)s(k) + z(k)\\
  & y_{\text{}}(k,\theta_{\text{}})  =   \sqrt{ P \alpha_\text{}} \mathbf{h}_\text{}^*(\theta_{\text{}})\mathbf{f}(k)s(k)  + z(k) \\      
  &  =   \sqrt{ N_\text{R} P \alpha_\text{}} g\mathbf{a}_\text{}^*(\theta_{\text{}})\mathbf{f}(k)s(k)  + z(k) \\       
 & =   \sqrt{\frac{N_\text{R} P \alpha_\text{}}{{N_\text{T}}}} g_\text{} s(k) \bigg( \sum_{m \in \mathcal{I}_M(k)} e^{-j\left(\frac{N_\text{T}-1}{2}-m\right)  \frac{2\pi d}{\lambda}  \cos(\theta_\text{R}) }  \times  \\ \nonumber & e^{j \left(\frac{N_\text{T}-1}{2}-m\right)  \frac{2\pi d}{\lambda}  \cos(\theta) }  +   \sum_{n \in \mathcal{E}_L(k)} e^{-j \left(\frac{N_\text{T}-1}{2}-n\right)  \frac{2\pi d}{\lambda}  \cos(\theta_\text{R}) }  \times \\ & e^{j \left(\frac{N_\text{T}-1}{2}-n\right)  \frac{2\pi d}{\lambda}  \cos(\theta) } -  \sum_{n \in \mathcal{O}_L(k)} e^{-j \left(\frac{N_\text{T}-1}{2}-n\right)  \frac{2\pi d}{\lambda}  \cos(\theta_\text{R}) }  \times \\ & e^{j \left(\frac{N-1}{2}-n\right)  \frac{2\pi d}{\lambda}  \cos(\theta) }\bigg)  + z (k)
\end{aligned}
\end{eqnarray}      
\begin{eqnarray}\label{arnl}
\vspace{-20mm}
\hspace{-22mm}= \underbrace{\sqrt{P \alpha} g}_{\substack{\text{effective }\\\text{channel gain}}} \underbrace{\sqrt{N_\text{R}}  \beta}_{\substack{\text{array gain }\\\text{}}} \underbrace{s(k)}_{\substack{\text{ information }\\\text{symbol}}}   +  \underbrace{{z}(k)}_{\substack{\text{additive }\\\text{noise}}},
\end{eqnarray}
where 
\begin{eqnarray}
\nonumber  \beta &=& \sqrt{\frac{1}{{N_\text{T}}}} \bigg (\sum_{m\in \mathcal{I}_M(k)} e^{j \left(\frac{N_\text{T}-1}{2}-m\right)  \frac{2\pi d}{\lambda} (\cos(\theta) - \cos(\theta_\text{R}))} \\ \nonumber  &+&   \sum_{n \in \mathcal{E}_L(k)}  e^{j \left(\frac{N_\text{T}-1}{2}-n\right)  \frac{2\pi d}{\lambda}  (\cos(\theta) -\cos(\theta_\text{R})) }   - \\ \label{anb}  &&  \sum_{n \in \mathcal{O}_L(k)} e^{j \left(\frac{N_\text{T}-1}{2}-n\right)  \frac{2\pi d}{\lambda} (\cos(\theta) -\cos(\theta_\text{R})) } \bigg).
\end{eqnarray}
Note when  $\theta = \theta_\text{R}$, the term $\beta = {\frac{M}{\sqrt{N_\text{T}}}}$ and the array gain in (\ref{arnl}) becomes a constant. When $\theta \not= \theta_\text{R}$, the term $\beta$ becomes a random variable. This randomizes the array gain at directions $\theta\not = \theta_\text{R}$, and as a result, jams eavesdroppers at these directions. In the following Lemma, we characterize the random variable $\beta$ for large number of transmit antennas $N_\text{T}$.
\begin {lemma} \label{l1}
For sufficiently large $N_{\text{T}}$ and $M$, and $\theta_\text{} \not = \theta_\text{R}$, $\beta$ converges to a complex Gaussian random variable with mean
\begin{eqnarray}\label{l1m1}
\mathbb{E}[ \beta ]= {\frac{{M}}{N_{\text{T}}\sqrt{N_{\text{T}}}}}\frac{ \sin\left( {N_{\text{T}}} \frac{\pi d}{\lambda}(\cos (\theta_{\text{}}) -\cos (\theta_{\text{R}}))  \right)} {\sin\left(  \frac{\pi d}{\lambda}(\cos (\theta_{\text{}}) -\cos (\theta_{\text{R}}))  \right)},
\end{eqnarray}  
and variance
 \begin{eqnarray}\label{l1v1}
 \text{var} [\beta] = \frac{N_{\text{T}}^2-M^2}{N^2_{\text{T}}}.
\end{eqnarray}  
\end{lemma}
\begin{proof}
See Appendix A for proof.
\end {proof}

		\begin{figure}
		\begin{center}
\includegraphics[width=3.5in]{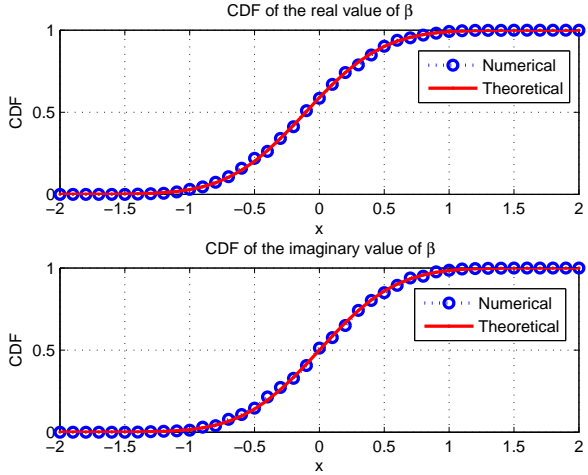}
\caption{Comparison of the numerical and theoretical values for the CDF of $\beta$ $F_{\beta}(x) = 1-Q\left( \frac{x-\mathbb{E}[\beta]}    {\sqrt{\text{var}[\beta]}}  \right) $; $N_\text{T}=64$, $M=48$, $\theta_\text{R}=100^{\circ}$, and $\theta_\text{}=60^\circ$.}\label{fig:cdc}
		\end{center}
	\end{figure}

In Fig. \ref{fig:cdc}, we plot the real and imaginary cumulative distribution function (CDF) of $\beta$. For validation purposes, we also plot the theoretical CDF of a Gaussian random variable with similar mean and variance. From the figure, we observe that numerical and theoretical (using (\ref{l1v1}), (\ref{eb301}), and (\ref{eib0})) CDFs are identical.

\section{Hybrid Beamforming with Opportunistic Noise Injection} \label{sec:ST2}

The proposed PHY security technique in the previous section enhances the communication link secrecy by randomizing the sidelobes of the far field radiation pattern. This jams the communication link in all non-receiver directions. It is sometimes desirable to radiate interference (or artificial noise) in controlled directions only. For example, if the vehicle is performing joint radar and communication \cite{preeti}, \cite{radar2i}, the jamming signal might interfere with the radar signal. In addition, the geometric design of the road restricts the location of potential eavesdroppers. Equipped with this prior information, it is sensible to jam receivers in threat regions only.  This allows more power to be radiated in threat regions and, as a result, increases the interference power and minimizes interference to other non-threat regions. 

\begin{figure}[t]
		\begin{center}
\includegraphics[width=2.8in]{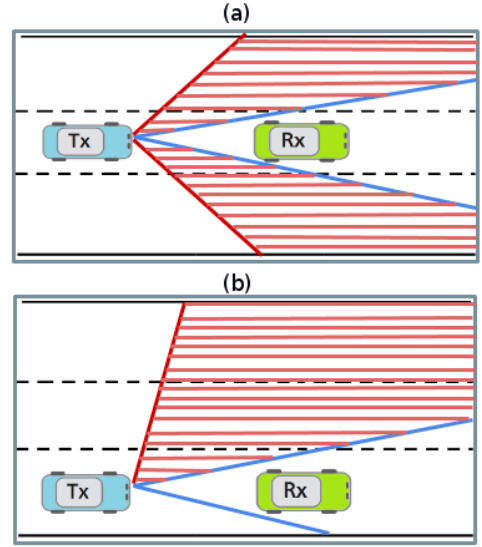}
				\caption{An example of two transmitters (in (a) and (b)) opportunistically injecting artificial noise in non-receiver directions. Regions with horizontal lines receive controlled noise.}
			\label{fig:car2}
		\end{center}
	\end{figure}

In this section, we propose a PHY security technique  for mmWave vehicular systems that exploits prior knowledge of the location of the target vehicle and road network geometry. We use side information of the road and lane width, traffic barriers, and curvature, to  opportunistically inject controlled artificial noise in potential threat areas (see Fig. \ref{fig:car2}).  In the following, we first introduce the proposed PHY security technique  using fully digital transmission precoders. Following that, we adopt the antenna architecture shown in Fig. \ref{fig:model} and design hybrid analog/digital transmission precoders that approximate fully digital transmission precoders with  $N_\text{RF}\ll N_\text{T}$ RF chains.

\subsection{Digital Precoder Design}
Let $\mathbf{f}_\text{s}$ be a data transmission beamformer and $\mathbf{f}_\text{n}$ be an artificial noise beamformer. Also let the set $\mathcal{T}$ define the range of angles along the road where potential eavesdroppers can be located. This can be obtained from prior knowledge of the road geometry.  The beamformer $\mathbf{f}_\text{s}$ is designed to maximize the beamforming gain at $\theta = \theta_\text{R}$. Under  our  LoS  and  ULA assumptions, the  beamforming  solution  that maximizes  the beamforming gain at the receiver is the  spatial  matched  filter \cite{trees}. Therefore $\mathbf{f}_\text{s}$ becomes
\begin{eqnarray}\label{dfs}
 {\mathbf{f}}_{\text{s}}  = \frac{1}{\sqrt{N_\text{T}}}\mathbf{a}(\theta_\text{R}).
 \end{eqnarray}

%Let $\mathbf{g} = [ g(\theta_1) \quad  ... \quad g(\theta_{L_\text{d}})]^{\mathrm{T}}$ be an $L_\text{d} \times 1$ vector and $L_\text{d} > N_\text{T}$ be the total number of possible transmit directions.
%A. Sayed, Fundamentals of Adaptive Filtering .   Hoboken, NJ, USA: Wiley, 2003.

To design the beamformer $\mathbf{f}_\text{n}$, we first obtain a matrix $\mathbf{B}$ of size $N_\text{T} \times N_\text{T}$ with $N_\text{T}$ columns orthogonal to the receiver's channel. Using a combination of these columns, we then design a unit norm beamformer $\mathbf{f}_\text{n}$ to have a constant projection on $\theta \in \mathcal{T}$, i.e. 
\begin{eqnarray}\label{g1}
\mathbf{a}^*(\theta)\mathbf{f}_\text{n} = q(\theta), 
 \end{eqnarray}
where $q(\theta) = 1 \text{ if } \theta \in \mathcal{T} \text { and } 0  \text{ otherwise.}$ Using the Householder transformation, the matrix $\mathbf{B}$ can be expressed as
\[
\mathbf{B} = \mathbf{I} - \frac{1}{N_\text{T}}\mathbf{a}(\theta_\text{R})\mathbf{a}^*(\theta_\text{R}),
\]
where $\mathbf{I}$ is the identity matrix. Since the columns of $\mathbf{B}$ are orthogonal to $\mathbf{a}^*(\theta_\text{R})$, their combination, and hence $\mathbf{f}_\text{n}$, will be orthogonal to $\mathbf{a}^*(\theta_\text{R})$ as well. Using $\mathbf{B}$, the design objective of (\ref{g1}) becomes finding a solution for 
\begin{eqnarray} \label{g2}
\mathbf{A}^*\mathbf{Bx} = \mathbf{q},
%\mathbf{A}^\mathbf{f}_\text{n} = \mathbf{g}
 \end{eqnarray} 
where $\mathbf{A}=[\mathbf{a}(\theta_1),...,\mathbf{a}(\theta_{L_\text{d}})]$ is the array response matrix, $L_\text{d} > N_\text{T}$ is the total number of possible transmit directions, $\mathbf{x}\in \mathcal{C}^{N_\text{T}}$, and $\mathbf{q} = [ q(\theta_1) \quad  ... \quad q(\theta_{L_\text{d}})]^{\mathrm{T}}$. The indices of the non-zero entries of $\mathbf{x}$ in (\ref{g2}) correspond to the selected columns of the matrix $\mathbf{B}$, and their values correspond to appropriate weights associated with each column. The least-squares estimate of  $\mathbf{x}$ is
\begin{eqnarray} \label{g3}
\hat{\mathbf{x}} = (\mathbf{Z}^*\mathbf{Z})^{-1}\mathbf{Z}^*\mathbf{q},
 \end{eqnarray} 
where $\mathbf{Z} = \mathbf{A^*B}$. Using (\ref{g1})-(\ref{g3}), the beamformer $\mathbf{f}_\text{n}$ becomes $\mathbf{f}_\text{n} = C\mathbf{B}\hat{\mathbf{x}}$, where  $C$ is a normalization constant that results in $\| {\mathbf{f}}_{\text{n}}\|^2 = 1$.

Based on $\mathbf{f}_\text{s}$ and  $\mathbf{f}_\text{n}$, the precoder $\mathbf{f}(k)$ becomes
\begin{eqnarray} \label{e15q}
\mathbf{f}(k) = \sqrt{\epsilon}\mathbf{f}_\text{s} + \sqrt{(1-\epsilon)}\mathbf{f}_\text{n} \eta(k),
 \end{eqnarray}
where $\epsilon $ denotes the power fraction allocated for data transmission and it takes a value between 0 and 1, and $\eta(k)$ is the artificial noise term.
The artificial noise term $\eta(k)$ can take any distribution, however, for simplicity of the analysis, we assume  $\eta(k)$ = $e^{j\Theta(k)}$, with $\Theta(k)$  uniformly distributed between 0 and $2\pi$. Using (\ref{e15q}), the received signal at an arbitrary receiver becomes
%\begin{eqnarray}  
%y_{\text{}}(k,\theta_{\text{}})   &=&  \sqrt{P\alpha_\text{} N_\text{R}} \mathbf{h}^*(\theta_{\text{}})\mathbf{f}(k){s}(k)+z_{\text{}}(k)\\   \label{e15} &=&
% \underbrace{\sqrt{P\alpha_\text{} \gamma N_\text{R}} \mathbf{h}^*(\theta_{\text{}})\mathbf{f}_\text{s}}_{\substack{\text{effective }\\\text{channel gain}}} \underbrace{{s}(k)}_{\substack{\text{information }\\\text{symbol}}} + \underbrace{\sqrt{P \alpha_\text{} (1-\gamma)N_\text{R} } \mathbf{h}^*(\theta_{\text{}})\mathbf{f}_\text{n}\eta(k)s(k)}_{\substack{\text{artifical }\\\text{noise}}}+\underbrace{z_{\text{}}(k)}_{\substack{\text{additive }\\\text{noise}}}.
% \end{eqnarray}
% 
% 
\begin{eqnarray}  
\hspace{-2mm} y_{\text{}}(k,\theta_{\text{}})   &=&  \sqrt{P\alpha_\text{} N_\text{R}} \mathbf{h}^*(\theta_{\text{}})\mathbf{f}(k){s}(k)+z_{\text{}}(k)\\  \nonumber  &=&
\underbrace{{s}(k)}_{\substack{\text{information }\\\text{symbol}}} (\underbrace{\sqrt{P\alpha_\text{} \epsilon N_\text{R}} g\mathbf{a}^*(\theta_{\text{}})\mathbf{f}_\text{s}}_{\substack{\text{effective }\\\text{channel gain}}} +\\ \label{e15} &&  \underbrace{\sqrt{P \alpha_\text{} (1-\epsilon)N_\text{R} } g\mathbf{a}^*(\theta_{\text{}})\mathbf{f}_\text{n}\eta(k)}_{\substack{\text{artifical }\\\text{noise}}}) +\underbrace{z_{\text{}}(k)}_{\substack{\text{additive }\\\text{noise}}}.
 \end{eqnarray} 
Note that $\mathbf{f}_\text{n}$ in (\ref{e15}) is orthogonal to $\mathbf{a}^*(\theta_{\text{R}})$, and therefore, the target receiver (at $\theta=\theta_{\text{R}}$ ) is not affected by the artificial noise term.

\subsection{Hybrid Analog/Digital Precoder Design}	
	Due to hardware constraints, the transmitter may not be able to apply the unconstrained entries of ${\mathbf{f}}_{\text{}}(k)$ in (\ref{e15q}) to form its beam. As discussed in \cite{ahmed1} and \cite{omar}, the number of RF chains $N_\text{RF}\ll N_\text{T}$, and the analog phase shifters have constant modulus and are usually quantized to yield $L<L_\text{d}$ constrained directions \cite{ahmed1}. One possible solution is to use a limited number of RF chains together with the quantized phase-shifters to form a hybrid analog/digital design. To design the hybrid precoder, we set $\mathbf{f}_\text{h}(k) =  \mathbf{F}_{\text{RF}}(k)\mathbf{f}_{\text{BB}}(k)$, where the subscript h refers to hybrid precoding, $\mathbf{F}_{\text{RF}}(k)$ is an $N_\text{T} \times N_{\text{RF}}$ precoder,  and $\mathbf{f}_{\text{BB}}(k)$  is an $N_{\text{RF}} \times 1$ digital (baseband) precoder. Consequently, the design of the precoder is accomplished by solving \cite{ahmed1}
	\begin{eqnarray}\label{p1}
	&&\hspace{-5mm} \{ \mathbf{F}^\star_{\text{RF}}(k), \mathbf{f}^\star_{\text{BB}}(k)   \} = \arg \min \|{\mathbf{f}}_{\text{}}(k) - \mathbf{F}_{\text{RF}}(k) \mathbf{f}_{\text{BB}}(k) \|_{\text{F}}, \\
	\nonumber && \hspace{-5mm}\text{s.t. }  [\mathbf{F}_{\text{RF}}(k)]_{:,i} \in \{ [\mathbf{A}_\text{can}]_{:,\ell}| 1\leq \ell \leq L  \}, i=1,\cdots, N_{\text{RF}}-1,\\
	\nonumber && \quad \|\mathbf{F}_{\text{RF}}(k) \mathbf{f}_{\text{BB}}(k)\|_{{F}}^2=1,
	\end{eqnarray}
where $\mathbf{A}_\text{can} = [\mathbf{a}(\theta_1),....,\mathbf{a}(\theta_L)]$  is an $N_{\text{}} \times L$ matrix that carries all set of possible analog beamforming vectors due to the constant modulus and angle quantization constraint on the phase shifters, and $\theta_l \in \{0, \frac{2\pi}{L}, ...,  \frac{2\pi (L-1)}{L} \}$ \cite{ahmed1}. Given the matrix of possible RF beamforming vectors  $\mathbf{A}_\text{can}$, the optimization problem in (\ref{p1}) can be reformulated as a sparse approximation problem which is solved using matching pursuit algorithms as proposed in \cite{omar} { {(Algorithm 1)}} to find $\mathbf{f}^\star_{\text{BB}}(k) $ and $ \mathbf{F}^\star_{\text{RF}}(k)$ and consequently $	{\mathbf{f}}_\text{h}(k)$  as follows
	\begin{eqnarray}\label{approx1}
	{\mathbf{f}}_\text{h}(k) =  \mathbf{F}^\star_{\text{RF}}(k) \mathbf{f}^\star_{\text{BB}}(k).
	\end{eqnarray}
	
One drawback of this solution is that the entries of the matrix $\mathbf{A}_\text{can}$  are constrained by the hardware limitations of the analog phase-shifters. This results in a grid mismatch since the unconstrained digital precoder is designed using a matrix $\mathbf{A}$ with unconstrained entires, i.e. $\theta_l$ is not constrained. This grid mismatch results in a spectral spillover and destroys the sparsity of the optimization problem in (\ref{p1}). This issue was resolved in \cite{ahmed1} and \cite{omar} by increasing the number of RF chains. While larger number of RF chains might be justifiable for cellular systems, the high power consumption and cost of these arrays might limit their use to luxurious vehicles only. In the following, we propose a simple technique that relaxes the RF limitations and as a result, requires less  RF chains when compared to the design proposed in \cite{ahmed1} and \cite{omar}.

 	\begin{figure*}[t]
 		\begin{center}
 %\hspace{-10mm} \includegraphics[width=8in,height=3in]{figs1/bf11.eps}
 \hspace{-15mm} \includegraphics[width=7in]{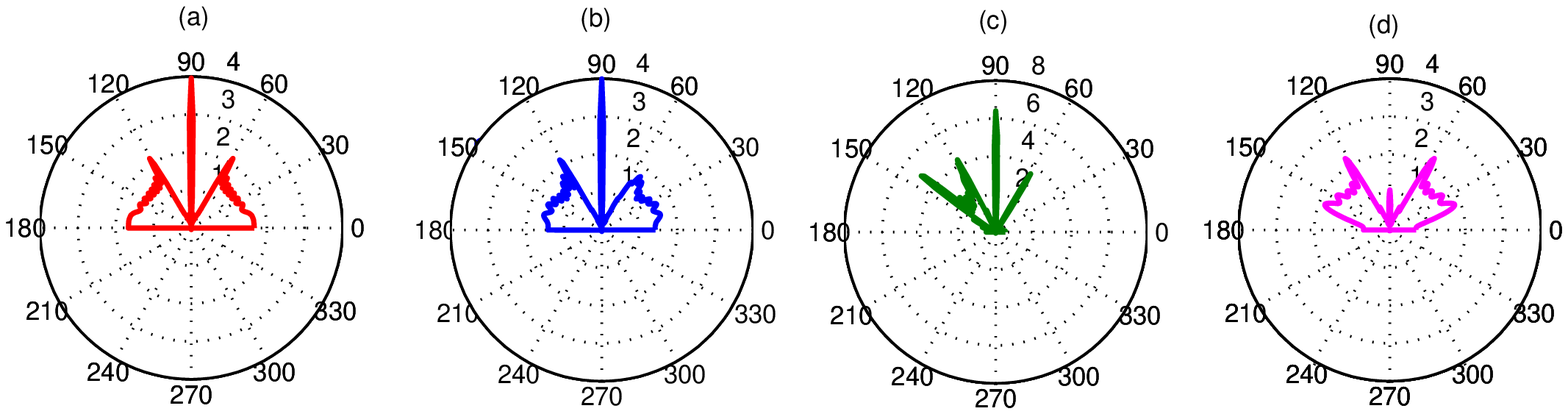}
\caption{Resulting beam pattern of  a patch antenna  with $N_\text{T}=32$ and a front-to-back ratio of 21.63 dB; (a) fully digital architecture, (b) proposed hybrid architecture with $N_\text{g}=8$, (c) hybrid architecture \cite{ahmed1}, \cite{omar}, (d) hybrid architecture using antenna switches \cite{sw}. The number of RF chains is fixed to $N_\text{RF} = 8$ for all hybrid architectures, 
$L_\text{d}=360$, $L=256$ (or 8-bit angle quantization), $\theta_\text{R}=90^\circ$, $\mathcal{T}\in \{ 0,..,60, 120,...,180 \}$, and $\epsilon = 0.1$.}
 			\label{fig:bf1}
 		\end{center}
 	\end{figure*}

%When  the ULA is steered at $-\theta$, its symmetry causes another main beam at ?'. The
%beam pointing at ?' can be suppressed by using a ground plane to absorb it or by using directional elements.

\subsection{Hybrid Design with Antenna Selection}	
In the previous hybrid design, the antenna spacing $d$ is kept constant. Therefore, the number of columns of the matrix $\mathbf{A}_\text{can}$ in (\ref{p1}) was $L$. When antennas are allowed to switch on/off, the number of columns becomes $(2^{N_\text{T}}-1)L$, since for every column, there are $2^{N_\text{T}}-1$ combinations. This provides additional degrees of freedom which could be used to reduce the number of RF chains at the expense of increased  computational complexity since the number of antennas is generally large in mmWave systems. To reduce the computational complexity, we randomly group $N_\text{g}$ antennas, and on/off switching is applied to each group instead of individual antennas. This reduces the number of columns of the of the matrix $\mathbf{A}_\text{can}$ to   $(2^{{N_\text{T}}/{N_\text{g}}}-1)L$.

 In Fig. \ref{fig:bf1}, we design the precoders $\mathbf{f}(k)$ and $\mathbf{f}_\text{h}(k)$ with $\theta_\text{R}=90^\circ$ and $\mathcal{T} \in \{0,...,60, 120,..,150\}$ degrees, and plot the resulting beam pattern, i.e. we plot $|\mathbf{a}^*(\theta)\mathbf{f}(k)|^2$ for the digital antenna architecture and $|\mathbf{a}^*(\theta)\mathbf{f}_\text{h}(k)|^2$ for the hybrid antenna architecture.  Fig. \ref{fig:bf1}(a) shows the pattern obtained when using a fully digital architecture. Fig. \ref{fig:bf1}(b) shows the pattern obtained when using the proposed hybrid architecture with antenna switching. As shown, the resulting pattern is similar to the pattern obtained when using a fully digital architecture with just 8 RF chains. Fig. \ref{fig:bf1}(c), shows the pattern obtained when using the hybrid architecture proposed in \cite{ahmed1} and \cite{omar}.  As shown, the resulting pattern is not accurate and requires more RF chains to realize good beam patterns. Fig. \ref{fig:bf1}(d)  shows the pattern obtained when using the hybrid architecture proposed in \cite{sw} with switches only. While some similarities exist between the resulting pattern and the fully digital pattern, we observe tremendous beamforming gain loss, especially at $\theta = \theta_\text{R}$. The reason for this is that the design in \cite{sw} is based on antenna subset selection (without phase shifters) which results in a beamforming gain loss as shown in Fig. \ref{fig:bf1}(d). In the proposed design, both phase-shifters and switches are jointly used to approximate the digital precoder with just a few RF chains.

\section{Secrecy Evaluation}\label{sec:ana}

In this section, we evaluate the performance of the proposed mmWave secure transmission techniques in terms of the secrecy rate. In our analysis, the transmitter is assumed to know the angular location of the target receiver but not of the potential eavesdropper. Both the target receiver and the eavesdropper are assumed to be equipped with a matched filter receiver and have perfect knowledge of their channels at all times.

The secrecy rate $R$ is defined as the maximum transmission rate at which information can be communicated reliably and securely, and is given by
\begin{eqnarray}\label{re1}
R &=&  [\log_2(1+\gamma_{\text{R}}) -\log_2(1+\gamma_{\text{E}})]^+,
 \end{eqnarray}
where $\gamma_{\text{R}}$  is the signal-to-noise ratio (SNR) at the target receiver, $\gamma_{\text{E}}$ is the SNR at the eavesdropper, and $a^+$  denotes $\max\{ 0,a \}$. We assume an eavesdropper with a sensitive receiver and, for mathematical tractability, the gain $g_\text{}$ is assumed to be constant and set to $g_\text{} = 1$. As shown in \cite{vuthaj}, directional beamforming at both the transmitter and the receiver make the coherence time to become quite long. This justifies the assumption of a fixed $g$ at the target receiver and allows us to account for the effects of the artificial noise on the eavesdropper and derive the secrecy rate. %{  We also assume that all receivers employ matched filter beamforming and have perfect channel knowledge, i.e. both the target receiver and the eavesdropper perfectly know their angle of arrivals and complex channel gains.}

\subsection{Analog Beamforming Secrecy Rate}
 The SNR at the target receiver at $\theta = \theta_\text{R}$  is given by (see (\ref{arnl}))
 \begin{eqnarray}  \label{gr1}
\hspace{-5mm}\gamma_{\text{R}} \hspace{-1mm}&=& \frac{P \alpha N_\text{R} |g\beta|^2  }{ \sigma^2} = \frac{P\alpha N_\text{R}  M^2  }{N_\text{T} \sigma^2}.
 \end{eqnarray}  

%\subsection{Analog Beamforming Secrecy Rate}
%%\subsubsection{SNR at Target Receiver}
% The SNR at the target receiver at $\theta = \theta_\text{R}$  is given by (see (\ref{arnl}))
% \begin{eqnarray}
%\hspace{-5mm}\gamma_{\text{R}} \hspace{-1mm}&=&\hspace{-1mm} \frac{(\mathbb{E}[   \sqrt{N_\text{R} P \alpha} \mathbf{h}^*(\theta_{\text{R}})\mathbf{f}(k) ])^2}{\text{var}[\sqrt{N_\text{R}P \alpha} \mathbf{h}^*(\theta_{\text{R}})\mathbf{f}(k)]+{\sigma^2} } \\ \label{gr1}
% \hspace{-1mm}&=&\hspace{-1mm} \frac{P \alpha N_\text{R}|g|^2 (\mathbb{E}[\beta])^2  }{P \alpha N_\text{R}|g|^2  \text{var}[\beta]+ \sigma^2} = \frac{2P\alpha N_\text{R}  M^2 \sin^2 ({\frac{\nu}{ D}}) }{N_\text{T} \sigma^2}.
% \end{eqnarray}  
%In (\ref{gr1}), the constant $|g|^2=\frac{1}{2} (1-e^{j\Phi_\text{R}})(1-e^{-j\Phi_\text{R}}) = 2\sin^2 (\frac{\Phi_\text{R}}{2})=2\sin^2 (\frac{\nu}{D})$, where $\nu=\frac{4\pi h_\text{t}h_\text{r}}{\lambda}$, and $\text{var}[ \beta]=0$ since $\beta = \frac{M}{\sqrt{N_\text{T}}}$ is a constant at $\theta=\theta_\text{R}$.

To derive the SNR at an eavesdropper at $\theta_\text{} \not = \theta_\text{R}$, we rewrite (\ref{arnl}) as
\begin{eqnarray}\label{arnl12}
% y_{\text{}}(k,\theta_{\text{}})= \underbrace{\sqrt{P \alpha} g}_{\substack{\text{effective }\\\text{channel gain}}} \underbrace{\sqrt{N_\text{R}}  \beta}_{\substack{\text{array gain }\\\text{}}} \underbrace{s(k)}_{\substack{\text{ information }\\\text{symbol}}}   +  \underbrace{{z}(k)}_{\substack{\text{additive }\\\text{noise}}},
 y_{\text{E}}(k,\theta_{\text{}})&=& \sqrt{P \alpha_\text{E} g_\text{E}N_\text{E}}  \beta  +  z_\text{E}(k)\\ \label{arnl13}
 &=& \sqrt{P \alpha_\text{E} g_\text{E}N_\text{E}}  (\beta_1 +\beta_2) s(k)  +  z_\text{E}(k),
\end{eqnarray}
where $N_\text{E}$ is the number of antennas at the eavesdropper, $g_\text{E}$ is the eavesdropper's channel gain, $\alpha_\text{E}$ is the eavesdropper's path loss, and $z_{\text{E}}(k) \sim  \mathcal {CN}(0,\sigma_\text{E}^2)$ is the additive noise at the eavesdropper. In (\ref{arnl12}), the constant $ \beta_1=\mathbf{E}[\beta]$ is the beamforming gain at the eavesdropper and the term $\beta_2=\beta-\mathbf{E}[\beta]$ is a zero mean random variable that represents the  artificial noise at the eavesdropper. Using (\ref{arnl13}) and Lemma \ref{l1}, the SNR  at the eavesdropper at $\theta_\text{} \not = \theta_\text{R}$ becomes
\begin{eqnarray} \label{snrem}
\gamma_{\text{E}}  = \frac{P \alpha_\text{E} N_\text{E} |g_\text{E}|^2 |\beta_1|^2  }{P \alpha_\text{E} N_\text{E} |g_\text{E}|^2 \text{var}[\beta_2] + \sigma_\text{E}^2},% =  \frac{|\mathbb{E}[\beta]|^2  }{ \text{var}[\beta]},  %and letting $N_\text{E} \rightarrow \infty$, 
\end{eqnarray} 
where the random variable $\beta$ is defined in (\ref{bbeta}). Letting $N_\text{E} \rightarrow \infty$ we obtain 
\begin{eqnarray} \label{snrem2}
\bar{\gamma}_{\text{E}} =  \frac{|\mathbb{E}[\beta]|^2  }{ \text{var}[\beta]},  %and letting $N_\text{E} \rightarrow \infty$, 
\end{eqnarray} 
where $\bar{\gamma}_{\text{E}}$ is the limit of $\gamma_\text{E}$ for large numbers of receive antennas, and $\gamma_\text{E}<\bar{\gamma}_{\text{E}} $. From Lemma \ref{l1},
\begin{eqnarray}\label{l1m1a}
\mathbb{E}[ \beta ] = {\frac{{M}}{N_{\text{T}}\sqrt{N_{\text{T}}}}}\frac{ \sin\left( {N_{\text{T}}} \frac{\pi d}{\lambda}(\cos (\theta_{\text{}}) -\cos (\theta_{\text{R}}))  \right)} {\sin\left(  \frac{\pi d}{\lambda}(\cos (\theta_{\text{}}) -\cos (\theta_{\text{R}}))  \right)},
\end{eqnarray}  
and
 \begin{eqnarray}\label{l1v1a}
 \text{var} [\beta] = \frac{N_{\text{T}}^2-M^2}{N^2_{\text{T}}}.
\end{eqnarray}
Using (\ref{l1m1a}) and (\ref{l1v1a}), (\ref{snrem2}) becomes
 \begin{eqnarray} \label{se}
  \bar{\gamma}_\text{E}  =  \frac{M^2}{N_{\text{T}}(N_{\text{T}}^2-M^2)} \bigg( \frac{ \sin^2\left( {N_{\text{T}}} \frac{\pi d}{\lambda}(\cos (\theta_{\text{}}) -\cos (\theta_{\text{R}}))  \right)} {\sin^2\left(  \frac{\pi d}{\lambda}(\cos (\theta_{\text{}}) -\cos (\theta_{\text{R}}))  \right)} \bigg).
\end{eqnarray}  %
%\subsubsection{Secrecy Rate}
Using (\ref{gr1}) and (\ref{se}), we quantify a lower bound on (\ref{re1}) as follows
%Based on $\bar{\gamma}_\text{E}$, we quantify a lower bound on (\ref{re1}) as follows
%Based on $\bar{\gamma}_\text{E} $, we quantify a lower bound on (\ref{re1}) based on this  To calculate the secrecy rate, we substitute (\ref{gr1}) and (\ref{se}) in (\ref{re1}) to obtain
\begin{eqnarray}\label{Rsb1}
\nonumber R\hspace{-3mm} &>& \hspace{-3mm}  \bigg[ \log_2\bigg(1+\frac{P\alpha N_\text{R}  M^2  }{N_\text{T} \sigma^2} \bigg) - \log_2\bigg(1+  \frac{M^2}{N_{\text{T}}(N_{\text{T}}^2-M^2)} \times  \\ \label{sar} && \hspace{-1 mm}    \frac{ \sin^2\left( {N_{\text{T}}} \frac{\pi d}{\lambda}(\cos (\theta_{\text{}}) -\cos (\theta_{\text{R}}))  \right)} {\sin^2\left(  \frac{\pi d}{\lambda}(\cos (\theta_{\text{}}) -\cos (\theta_{\text{R}}))  \right)} \bigg)   \bigg]^+. 
 \end{eqnarray} 
 Equation (\ref{Rsb1}) shows that the secrecy rate is a function of the subset size $M$. Increasing $M$ increases the rate at the target receiver (first part of (\ref{Rsb1})) at the expense of lower noise variance at $\theta \not = \theta_\text{R}$ as shown in the second part part of (\ref{Rsb1})). Therefore, there is an optimum value of $M$ that maximizes the secrecy rate in (\ref{Rsb1}). 
 
 \begin{remark}
 To guarantee secrecy, the value of $M$ is selected to satisfy $ \gamma_\text{R}>\bar{\gamma}_\text{E}$, i.e.
 \begin{eqnarray} \label{iem}
\frac{ \rho M^2  }{N_\text{T}} > \frac{M^2u(\theta)}{N_{\text{T}}(N_{\text{T}}^2-M^2)} ,
 \end{eqnarray} 
 where $\rho =\frac{P\alpha N_\text{R}}{\sigma^2} $ and $u(\theta)= \frac{ \sin^2\left( {N_{\text{T}}} \frac{\pi d}{\lambda}(\cos (\theta_{\text{}}) -\cos (\theta_{\text{R}}))  \right)} {\sin^2\left(  \frac{\pi d}{\lambda}(\cos (\theta_{\text{}}) -\cos (\theta_{\text{R}}))  \right)}$. Solving for $M$ we obtain the following upper-bound on $M$
\begin{eqnarray}\label{um}
M< \sqrt{N_{\text{T}}^2 - \frac{u(\theta)}{\rho}  }.
 \end{eqnarray} 
 From (\ref{um}) we observe that if $u(\theta)=0$  (eavesdropper is in one of the antenna nulls) or $\rho \rightarrow \infty$ ($\sigma^2 \rightarrow 0$), then $M=N_\text{T}-1$ satisfies the inequality in (\ref{iem}). We also observe that if $\sigma^2$ is high, i.e. $\rho$ is low, then the upper-bound in (\ref{um}) decreases and higher values of $M$ are required to increase the artificial noise at non-receiver directions. Note that this bound is pessimistic since we assumed an eavesdropper with infinite array gain.
 \end{remark}

\subsection{Hybrid Beamforming with Opportunistic Noise Injection Secrecy Rate}
To derive the SNR at the receiver and eavesdropper, we assume that there  are a sufficient number of RF chains to perfectly approximate the digital precoder.
 %\subsubsection{SNR at Target Receiver}
The receive SNR at the target receiver (at $\theta = \theta_\text{R}$) can be written as (see (\ref{e15}))
 \begin{eqnarray} \label{Hgr1}
 \gamma_{\text{R}} = \frac{ N_\text{R} P \alpha \epsilon  | \mathbf{h}^*(\theta_{\text{R}})\mathbf{f}_\text{s}|^2 }{\sigma^2 } = \frac{P \alpha \epsilon  N_\text{R}N_\text{T}  }{ \sigma^2_{{}} },
   \end{eqnarray}  
where the second term of (\ref{e15}) does not appear in (\ref{Hgr1}) since $\mathbf{f}_\text{n}$ is orthogonal to $\mathbf{h}^*(\theta_{\text{R}})$, and $ |\mathbf{h}^*(\theta_{\text{R}})\mathbf{f}_\text{s}|^2=  N_\text{T}$. 

The SNR at the eavesdropper (at $\theta \not = \theta_\text{R}$ ) can be derived as 
  \begin{eqnarray}
 \gamma_{\text{E}} \hspace{-2mm} &=& \hspace{-2mm} \frac{\epsilon N_\text{E} P \alpha_\text{E} |\mathbf{h}^*(\theta_{\text{}})\mathbf{f}_s|^2}{(1-\epsilon) N_\text{E} P \alpha_\text{E} |\mathbf{h}^*(\theta_{\text{}})\mathbf{f}_n|^2 \text{var}[\eta(k)]  +{\sigma_\text{E}^2} } \\ \label{e251} &=& \hspace{-2mm} \frac{P \alpha_\text{E}  \epsilon N_\text{E}   |g_{\text{E}}|^2  |\mathbf{a}^*(\theta_{\text{}})\mathbf{f}_\text{s}|^2} {P \alpha_\text{E} (1-\epsilon)N_\text{E}    |g_{\text{E}}|^2  |\mathbf{a}^*(\theta_{\text{}})\mathbf{f}_n|^2  \text{var}[\eta(k)] +{\sigma_{\text{E}}^2} }. % \\  &=&\frac{\mathbb{E}[  |\sqrt{P\alpha_\text{E} \gamma N_\text{T} N_\text{E}}g_\text{E} \Gamma(\theta_\text{})|^2 ]} {\mathbb{E}[  |\sqrt{P\alpha_\text{E} N_\text{E}}g_{\text{E}}\mathbf{a}^*(\theta_{\text{R}})\mathbf{f}_\text{n}(k)(1-\Gamma(\theta_\text{}))|^2 ]+{\sigma_{\text{E}}^2} } \\ \label{Hgr2} & =&  \frac{ {P\alpha_\text{E} \gamma N_\text{T} N_\text{E}} (\Gamma(\theta_\text{}))^2 }{P \alpha_\text{E} N_\text{E}(1-\Gamma(\theta))^2\mathbb{E}[|\mathbf{a}^*(\theta)\mathbf{f}_\text{n}(k)|^2]  +{\sigma_{\text{E}}^2}}.
%    and it is optimized for each value of $\mu(\mathcal{T})$
%The noise bearing signal $\mathbf{f}_\text{n}(k)$ in (\ref{Hgr2})  is designed such that it has a constant projection on the array response vectors $\mathbf{a}(\theta), \theta\in \mathcal{T}$ (see (\ref{d1})-(\ref{d3})) for every transmission symbol. Assuming that the noise bearing signal $\mathbf{f}_\text{n}(k)$ in (\ref{Hgr2}) has constant projection across $\theta\in \mathcal{T}$, the term $\mathbb{E}[|\mathbf{a}^*(\theta_{\text{}})\mathbf{f}_\text{n}(k)|^2] < \min (\frac{\pi}{\mu(\mathcal{T})},N_\text{T})(1-\gamma)$ \cite{love}, where $\mu(\mathcal{T})$ is the length of the angle interval in $\mathcal{T}$.  Letting $N_\text{E} \rightarrow \infty$ we obtain 
 \end{eqnarray}   
 Letting $N_\text{E} \rightarrow \infty$ we obtain 
   \begin{eqnarray}  \label{e25}
\bar{\gamma}_{\text{E}} = \frac{\epsilon  |\mathbf{a}^*(\theta_{\text{}})\mathbf{f}_\text{s}|^2} {(1-\epsilon) |\mathbf{a}^*(\theta_{\text{}})\mathbf{f}_n|^2 },
  \end{eqnarray}   
where ${\gamma}_{\text{E}}<\bar{\gamma}_{\text{E}}$ and $\text{var}[\eta(k)]=1$. The term $|\mathbf{a}^*(\theta_{\text{}})\mathbf{f}_\text{s}|$ in (\ref{e25}) is a constant and it can be derived as
\begin{eqnarray}  
\nonumber && \hspace{-8mm} |\mathbf{a}^*(\theta_{\text{}})\mathbf{f}_\text{s}| \hspace{-2mm} \\ \nonumber 
   &=&  \hspace{-2mm}\bigg | \frac{1}{\sqrt{N_\text{T}}} \sum_{n=0}^{N_\text{T}-1} e^{-j\left(\frac{N_\text{T}-1}{2}-n\right)  \frac{2\pi d}{\lambda}  \cos(\theta_\text{R}) }  e^{j \left(\frac{N_\text{T}-1}{2}-n\right)  \frac{2\pi d}{\lambda}  \cos(\theta) } \bigg| \\ \nonumber \label{t2e}
   &=& \bigg | \frac{1}{\sqrt{N_\text{T}}} \sum_{n=0}^{N_\text{T}-1} e^{j\left(\frac{N_\text{T}-1}{2}-n\right)  \frac{2\pi d}{\lambda}  (\cos(\theta)-\cos(\theta_\text{R})) }\bigg | \\ &=& \frac{ \sin\left( {N_{\text{T}}} \frac{\pi d}{\lambda}(\cos (\theta_{\text{}}) -\cos (\theta_{\text{R}}))  \right)} {{\sqrt{N_\text{T}}}\sin\left(  \frac{\pi d}{\lambda}(\cos (\theta_{\text{}}) -\cos (\theta_{\text{R}}))  \right)}.
   \end{eqnarray}  
Note that the term $|\mathbf{a}^*(\theta_{\text{}})\mathbf{f}_\text{n}|^2$  in (\ref{e25}) represents the beamforming gain at $\theta \in \mathcal{T}$ and it is given by $|\mathbf{a}^*(\theta_{\text{}})\mathbf{f}_\text{n}|^2 = \frac{\pi}{\mu(\mathcal{T})}$ \cite{love}, where $\mu(\mathcal{T})$ is the length of the angle interval in $\mathcal{T}$, i.e. sector size.  This expression applies to sectors formed using ideal beam patterns which do not overlap. Since beam patterns of this form are usually unrealizable, there will be some ``leaked'' power radiated at angles $\theta \not \in \mathcal{T}$. This reduces the beamforming gain at $\theta \in \mathcal{T}$. To account for this,  we set $|\mathbf{a}^*(\theta_{\text{}})\mathbf{f}_\text{n}|^2 = \frac{\pi c_0}{\mu(\mathcal{T})}$, where $0<c_0<1$ is a constant that accounts for beam pattern imperfections. Therefore (\ref{e25}) can be written as % the SNR at the eavesdropper (at $\theta \not = \theta_\text{R}$ and $N_\text{E} \rightarrow \infty$)
   \begin{eqnarray} \label{snrea}
  \bar{ \gamma}_{\text{E}}&=&  \frac{ {\mu(\mathcal{T})} \epsilon \sin^2\left( {N_{\text{T}}} \frac{\pi d}{\lambda}(\cos (\theta_{\text{}}) -\cos (\theta_{\text{R}}))  \right)} { { \pi(1-\epsilon)c_0} {N_{\text{T}}}\sin^2\left(  \frac{\pi d}{\lambda}(\cos (\theta_{\text{}}) -\cos (\theta_{\text{R}}))  \right)}.
  \end{eqnarray}  
  Using (\ref{Hgr1}) and (\ref{snrea}),  the secrecy rate is lower bounded by
\begin{eqnarray} 
&& \nonumber \hspace{-10mm}  R > \bigg[ \log_2\left(1+ \frac{P \alpha \epsilon  N_\text{T}N_\text{R}  }{ \sigma^2}\right)  - \log_2 \bigg(1+ \\  \label{srx} &&   \frac{ {\mu(\mathcal{T})} \epsilon \sin^2\left( {N_{\text{T}}} \frac{\pi d}{\lambda}(\cos (\theta_{\text{}}) -\cos (\theta_{\text{R}}))  \right)} { { (1-\epsilon)c_0\pi} {N_{\text{T}}}\sin^2\left(  \frac{\pi d}{\lambda}(\cos (\theta_{\text{}}) -\cos (\theta_{\text{R}}))  \right)}\bigg)\bigg]^+.
 \end{eqnarray}
Equation (\ref{srx}) shows that the secrecy rate is a function of the transmission power fraction $\epsilon$ and the sector size $\mu(\mathcal{T})$. Increasing $\epsilon$ results in a lower secrecy rate since more power will be invested in data transmission to the target receiver rather than artificial noise injection. We also observe that for fixed $\epsilon$, increasing $\mu(\mathcal{T})$ (second part of (\ref{srx})) decreases the secrecy rate since the artificial noise power will be radiated over larger sector sizes. Therefore, for a fixed transmission power, the transmitter can optimize the secrecy rate by choosing appropriate sector sizes. Prior road geometry information could be exploited to optimize the sector size $\mu(\mathcal{T})$.

 \begin{remark}
 To guarantee secrecy, the value of $\epsilon$ is selected to satisfy $ \gamma_\text{R}>\bar{\gamma}_\text{E}$, i.e.
 \begin{eqnarray} \label{ieme}
 \zeta \epsilon  > \frac{ {\mu(\mathcal{T})} \epsilon \nu(\theta) } { { \pi(1-\epsilon)c_0}},
 \end{eqnarray} 
 where $\zeta =\frac{P\alpha N_\text{T} N_\text{R}}{\sigma^2} $ and $\nu(\theta)= \frac{ \sin^2\left( {N_{\text{T}}} \frac{\pi d}{\lambda}(\cos (\theta_{\text{}}) -\cos (\theta_{\text{R}}))  \right)} {N_\text{T}\sin^2\left(  \frac{\pi d}{\lambda}(\cos (\theta_{\text{}}) -\cos (\theta_{\text{R}}))  \right)}$. Solving for $\epsilon$ we obtain the following upper-bound 
\begin{eqnarray}\label{umee}
\epsilon< 1-   \frac{\nu(\theta)\mu(\mathcal{T})}{\zeta\pi c_0}.
 \end{eqnarray}

 From (\ref{umee}) we observe that if $\nu(\theta)=0$ (i.e., eavesdropper is in one of the antenna nulls) or $\zeta \rightarrow \infty$ (i.e., $\sigma^2 \rightarrow 0$), then $\epsilon\approx 1 <1$ satisfies the inequality in (\ref{ieme}) and most transmission power can be directed towards symbol transmission. We also observe that if the noise variance at the target receiver is high, i.e. $\zeta$ is low, then the upper-bound in (\ref{umee}) decreases, and as a result, the sectoe size $\mu(\mathcal{T})$ should be reduced and/or more power power should be directed towards artificial noise injection at non-receiver directions. Note that the bound in  (\ref{umee}) is pessimistic since an eavesdropper with infinite array gain is assumed.
 \end{remark}

\section{Numerical Results and Discussions } \label{sec:num}
In this section, we study the performance of the proposed security techniques in the presence of a single eavesdropper with a sensitive receiver. The system operates at 60 GHz with a bandwidth of 50 MHz and an average transmit power of 37dBm. A standard two-ray log-distance path loss model with exponent 2 is used to model the mmWave communication channel.  Unless otherwise specified, the number of antennas at the target receiver is $N_\text{R} = 16$ and the number of antennas at the eavesdropper is $N_\text{E} = 500$. For simulations purposes, the eavesdropper's channel is assumed to be random with $g_\text{E}\sim \mathcal{CN}(0,1)$.  The distance from the transmitter to the receiver is set to 50 meters, the distance from the transmitter to the eavesdropper is set to 10 meters.

\begin{figure}[t]
		\begin{center}
\includegraphics[width=3.5in]{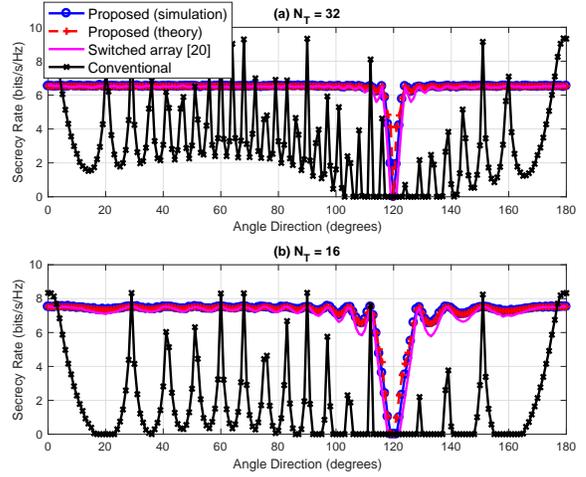}
				\caption{Comparison between the simulated and theoretical (eq. (\ref{sar})) secrecy rate for linear arrays with different length when steered to $\theta_\text{R}=120^{\circ}$; $M=12$.}
			\label{fig:rate_all}
		\end{center}
	\end{figure}

In Figs. \ref{fig:rate_all}(a) and (b), we plot the secrecy rate achieved by the proposed analog beamforming technique versus the eavesdropper's angular location for $N_\text{T} = 32$ and $N_\text{T} = 16$. For both cases, we observe that the secrecy rate of the proposed technique is high at all angular locations except at the target receiver's angular location ($\theta_\text{R}=120^\circ$).  We also observe that the switched array technique proposed in \cite{dm5} provides high secrecy rate, while conventional array techniques provide poor secrecy rates. The reason for this is that conventional array transmission techniques result in a constant radiation pattern at the eavesdropper while, the proposed and the switched array techniques randomize the radiation pattern at the eavesdropper, thereby jamming the eavesdropper. For the proposed and switched array transmission techniques, no randomness is experienced at the target receiver, and the secrecy rate is minimum when the eavesdropper is located in the same angle with the target receiver, which is generally not the case in practice. We also observe that the proposed technique is better than switched array techniques especially when the eavesdropper is closely located to the target receiver, i.e. when $|\theta_\text{R}-\theta|$ is small, which is typically the worst case scenario. When $|\theta_\text{R}-\theta|$ is large, the figure shows that the switched array technique proposed in \cite{dm5} provides comparable secrecy rates with the proposed technique. The proposed technique, however, is still superior to \cite{dm5} since: (i) The switched array technique requires antenna switches to jam eavesdroppers. The proposed technique does not require antenna switches. (ii) The idle antennas in the switched array technique creates a sparse array which could be exploited by adversaries to precancel the jamming signal. The proposed technique uses all antennas, therefore making it difficult to breach.  

In some instances,  Figs. \ref{fig:rate_all}(a) and (b) show that conventional array techniques provide  better secrecy rates when compared to the proposed and switched array techniques. This is particularly observed when the eavesdropper falls in one of the nulls (or closer to the null directions) of the transmit array pattern. Since conventional techniques use all antennas for data transmission, these techniques would result in higher secrecy rates when the eavesdropper falls in one of the nulls (or closer to the null directions) of the transmit array pattern.

%	\begin{figure}[t]
%		\begin{center}
%\includegraphics[width=3.65in]{rate_angle_Ne.eps}
%\caption{Comparison between the simulated and theoretical secrecy rate for a linear array steered to $\theta_\text{R}=120^{\circ}$ with different number of receiver antennas at the eavesdropper; $N_\text{T}=32$ and $M=12$.}
%			\label{fig:rate_all2}
%		\end{center}
%	\end{figure}

		\begin{figure}[t]
		\begin{center}
\includegraphics[width=3.5in]{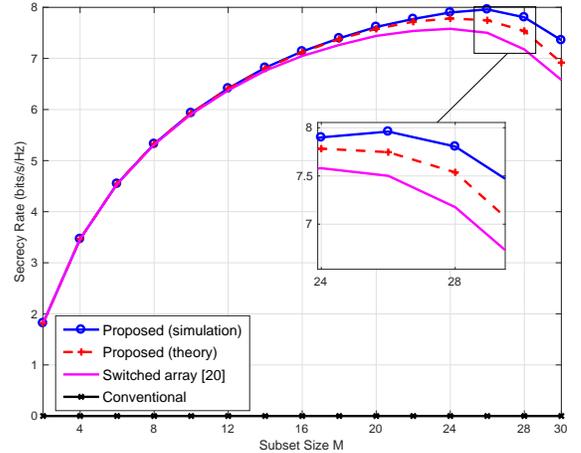}
				\caption{Simulated and theoretical secrecy rate average over the transmit angle $\theta$ versus the subset size $M$ for linear a array steered to $\theta_\text{R}=120^{\circ}$; $\theta \in \{\{110^\circ,...,130^\circ\}\setminus\{\theta_\text{R}\}\}$, $N_\text{T}=32$.}
			\label{fig:rate_sub}
		\end{center}
	\end{figure}

%In Figs. \ref{fig:rate_all2}(a) and (b), we plot the secrecy rate achieved by the proposed analog beamforming technique versus the eavesdropper's angular location for $N_\text{E} = 64$ and $N_\text{E} = 16$ antennas at the eavesdropper. Increasing the number of antennas at the eavesdropper results in lower secrecy rate when using conventional transmission techniques (without PHY security) due to increased receive SNR at the eavesdropper. The secrecy rate achieved by the proposed and the switched array techniques hardly changes with $N_\text{E}$. As the artificial noise created by these techniques increases with the array gain at the eavesdropper, increasing $N_\text{E}$ will hardly affect the secrecy rate.

In Fig. \ref{fig:rate_sub} we examine the impact of the transmission subset size $M$  on the secrecy rate of the analog beamforming technique. We observe that as the subset size $M$ increases, the secrecy rate of the proposed technique increases, plateaus, and then decreases. The reason for this is that as $M$ increases, more antennas are co-phased for data transmission. On one hand, this increases the beamforming gain at the target receiver. On the other hand, increasing $M$ decreases the variance of the artificial noise at the eavesdropper. Therefore, there is a trade-off between the beamforming gain at the receiver and the artificial noise  at a potential eavesdropper, and there exists an optimum value of $M$ that maximizes the secrecy rate. In Fig. \ref{fig:rate_sub}, we also show that the proposed technique achieves higher secrecy rate when compared to conventional transmission techniques and the switched array technique proposed in \cite{dm5}, especially for higher values of $M$.  Moreover, we show that the theoretical bound of the secrecy rate  is tight when compared to the secrecy rate obtained by Monte Carlo simulations. This verifies the correctness of our analysis and enables us to gain further insights into the impact of key parameters such as, the subset size $M$ and number of transmit/receive antennas, on the system performance.

		\begin{figure}[t]
		\begin{center}
\includegraphics[width=3.5in]{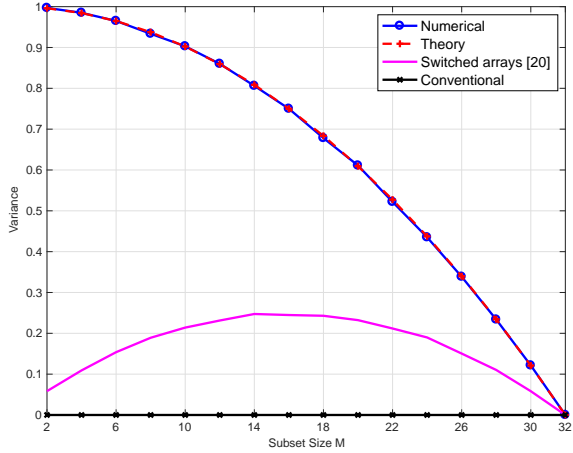}
				\caption{Numerical and theoretical (eq. (\ref{var5})) values of the beam pattern variance at $\theta = 115^{\circ}$ versus the subset size $M$ for linear a array steered to $\theta_\text{R}=120^{\circ}$; $N_\text{T}=32$.}
			\label{fig:var}
		\end{center}
	\end{figure}

			\begin{figure}[t]
			\begin{center}
	\includegraphics[width=3.5in]{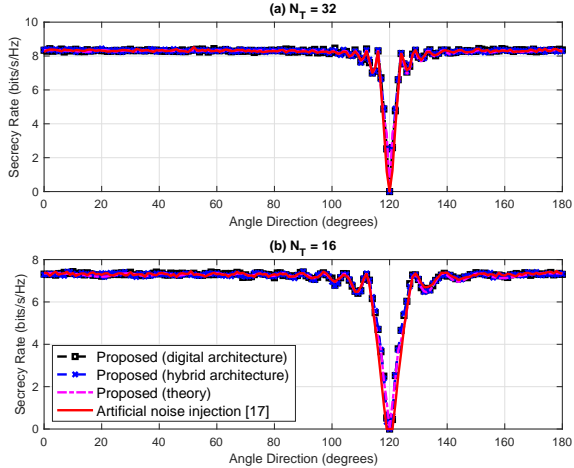}
\caption{ Numerical and theoretical (eq. (\ref{srx})) values of the secrecy rate versus the eavesdroppers angular direction for different values of $N_\text{T}$. Artificial noise is injected in all non-receiver directions with $\epsilon=0.5$. For the hybrid architecture, $N_\text{RF}$ = 10 with 6-bit angle quantization, and $N_\text{g} = 8$.}
				\label{fig:rate_theta}
			\end{center}
		\end{figure}

		\begin{figure}[t]
			\begin{center}
	\includegraphics[width=3.5in]{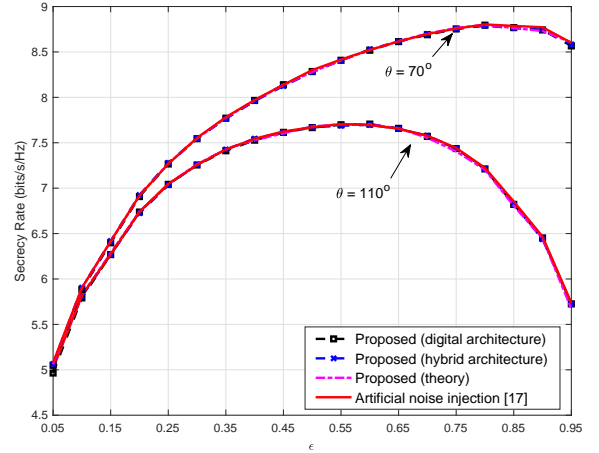}
\caption{ Secrecy rate versus the transmission power fraction $\epsilon$. Artificial noise is injected in all non-receiver directions, $N_\text{T}=32$, and $\theta_\text{R}=120^{\circ}$. For the hybrid architecture, $N_\text{RF} = 10$ with 6-bit angle quantization, and $N_\text{g}= 8$.}
				\label{fig:rate_gamma2}
			\end{center}
		\end{figure}

In Fig. \ref{fig:var} we plot the numerical and theoretical values of the beam pattern variance at $\theta = 115^{\circ}$ when using the proposed technique and the switched array technique. Note that higher beam pattern variance increases the variance of the artificial noise. As shown, the proposed technique provides higher beam pattern variance when compared to the switched array and conventional techniques. In the switched array technique, only $M$ antennas are used to generate the jamming signal, while the remaining $N_\text{T}-M$ antennas are idle. The proposed technique uses $M$ antennas for data transmission and  $N_\text{T}-M$ antennas to randomize the beam pattern at potential eavesdroppers. This results in higher secrecy rate. The variance of the conventional technique is zero since the beam pattern is fixed and the transmitter does not generate any artificial noise.

			\begin{figure}[t]
			\begin{center}
\hspace{-8mm}	\includegraphics[width=3.8in]{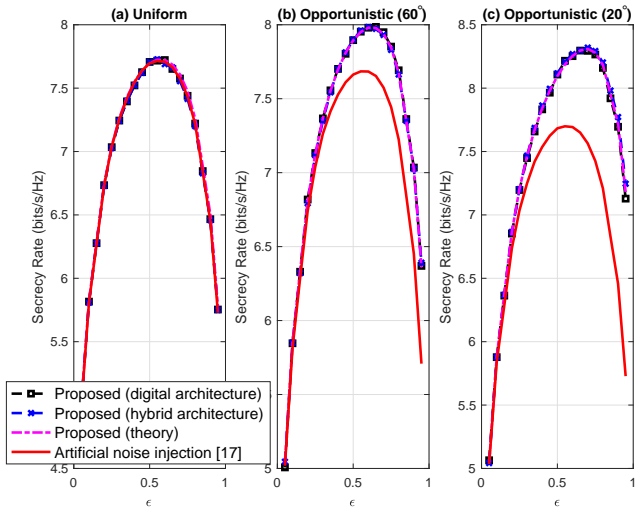}
\caption{Secrecy rate versus the transmission power fraction $\epsilon$. The number of antennas is fixed to $N_\text{T}=32$ with 8-bit angle quantization, $\theta_\text{R}= 120^{\circ}$, and $\theta_\text{}= 110^{\circ}$. For the hybrid architecture, $N_\text{RF} = 10$, and $N_\text{g} = 8$. In (a), artificial noise is uniformly injected in all non-receiver directions, in (b) artificial noise is opportunistically injected $30^\circ$ left and $30^\circ$ right to the receiver, and (c)  artificial noise is opportunistically injected $15^\circ$ left and $15^\circ$ right to the receiver.}
				\label{fig:rate_gamma}
				%C_o=0.6
			\end{center}
		\end{figure}

In Figs.\ref{fig:rate_theta}(a) and (b), we plot the secrecy rate versus the eavesdroppers angular direction when implementing the proposed security technique with opportunistic noise injection using a digital architecture and the proposed hybrid architecture. For comparison, we also plot the secrecy rate achieved when implementing the security technique proposed in \cite{anr1} which injects artificial noise in the null space of the target receiver. From the figures we observe that the secrecy rate of the proposed technique is  high at all angular locations except at the target receiver's angular location ($\theta_\text{R}= 120^\circ)$. We also observe that the secrecy rate (theoretical and simulation) achieved when using the proposed hybrid architecture, with  10 RF chains, is similar to the secrecy rate achieved when using a fully digital architecture and  when implementing the security technique in \cite{anr1}. This reduced number of RF chains reduces the complexity and cost of the array and makes the proposed architecture a favorable choice for mmWave vehicular applications.

In Fig. \ref{fig:rate_gamma2}, we plot the secrecy rate versus the transmission power fraction $\epsilon$ when implementing the proposed security technique and the technique in \cite{anr1}. The figure shows that more noise power (lower value of $\epsilon$) is required to maximize the secrecy rate if the eavesdropper is adjacent to the target receiver, while less noise power is required to maximize the secrecy rate when the eavesdropper is not adjacent to the target receiver. The reason for this is that the side lobe gain is typically higher for angles closer to the main beam (at $\theta_\text{R}$). This gain is lower for angles that are further from the main beam. Therefore, for the uniform noise injection case,  the value of $\epsilon$ can be opportunistically set to maximize the secrecy rate by taking into account possible threat regions in vehicular environments.

Finally in Figs. \ref{fig:rate_gamma}(a)-(c), we study the impact of the transmission power fraction $\epsilon$ on the secrecy rate. In all figures we observe that the secrecy rate is a concave function of $\epsilon$, and there is an optimal $\epsilon$ that maximizes the secrecy rate. The reason for this is that as $\epsilon$ increases, more power in utilized for symbol transmission rather than artificial noise injection. This results in an SNR improvement at the eavesdropper and a hit in the secrecy rate. Therefore, there is an optimum value of $\epsilon$ that maximizes the secrecy rate. Fig. \ref{fig:rate_gamma}(a) shows that the secrecy rate achieved by the proposed security technique when injecting artificial noise omni-directionally (excluding the target receiver's direction)  is similar to that achieved by the security technique proposed in \cite{anr1}. The reason for this is that in omni-directional transmission, the array gain can not be exploited to enhance the artificial noise. Nonetheless, the proposed technique achieves a similar rate with 10 RF chains when compared to \cite{anr1} that requires a fully digital architecture. Figs. \ref{fig:rate_gamma}(b) and (c) show the secrecy rate achieved by the proposed security technique when opportunistically  injecting artificial noise in  $60^\circ$ and $30^\circ$ sectors in the direction of the receiver (excluding  the angle $\theta_\text{R}$). The figures show that the secrecy rate of the proposed technique increases when artificial noise is injected in predefined sectors. The reason for this increase is that more power can be directed towards threat directions rather than simply spreading the power across all directions. Another reason is that the proposed technique exploits the multiple antennas to achieve a beamforming gain in the directions of interest, instead of injecting the noise power in all non-receiver (or orthogonal) directions as done in \cite{anr1}. This results in a higher noise variance, and as a result, improves the secrecy rate.

% % % % % % % % Insert analog figures here

%
%%%%%%%%%%%%%%%%%%%%%%%%%%%%%%%%%%%%%%%%%%%%%%%%%%%%%%%%%%%%%%%%%%%%%%%%%%%%%%%%%%%%%%%%%%%%%%%%%%%%%%%%%%%%%
\section{Conclusion} \label{sec:con}
%%%%%%%%%%%%%%%%%%%%%%%%%%%%%%%%%%%%%%%%%%%%%%%%%%%%%%%%%%%%%%%%%%%%%%%%%%%%%%%%%%%%%%%%%%%%%%%%%%%%%%%%%%%%%

In this paper,  we proposed two PHY security techniques for  mmWave vehicular communication systems. Both techniques respect mmWave hardware constraints and take advantage of the large antenna arrays at the mmWave frequencies to jam potential eavesdropper with sensitive receivers. This enhances the security of the data communication link between the  transmitter and the target receiver. To reduce the complexity and cost of fully digital antenna architectures, we proposed a new hybrid transceiver architecture for mmWave vehicular systems. We also introduced the concept of opportunistic noise injection and showed that opportunistic noise injection improves the secrecy rate when compared to uniform noise injection.  The proposed security techniques are shown to achieve secrecy rates close to that obtained by techniques that require fully digital architectures with much lower number of RF chains and without the need for the exchange of encryption/decryption keys. This makes the proposed security techniques favorable for time-critical road safety applications and vehicular communication in general.

\section*{Appendix A} 
The random variable $\beta$ in (\ref{anb}) can be rewritten as
\begin{eqnarray} \label{anba1} 
\nonumber \beta &=& \sqrt{\frac{1}{{N_\text{T}}}} \bigg (\sum_{m\in \mathcal{Z}(k)} e^{j \left(\frac{N_\text{T}-1}{2}-m\right)  \frac{2\pi d}{\lambda} (\cos(\theta) -\cos(\theta_\text{R}))} \\  \label{anba1}  &-&  \sum_{n \in \mathcal{O}_L(k)} e^{j \left(\frac{N_\text{T}-1}{2}-n\right)  \frac{2\pi d}{\lambda}  (\cos(\theta) -\cos(\theta_\text{R})) } \bigg),
\end{eqnarray}
where the set $\mathcal{Z}(k) =  \{\mathcal{I}_M(k) \cup \mathcal{E}_L(k) \}$, $\theta\not = \theta_\text{R}$, the cardinality of the set $\mathcal{Z}(k)$ is $ M+\frac{N_\text{T}-M}{2}$ and the cardinality of the set $\mathcal{O}_L(k)$ is $\frac{N_\text{T}-M}{2}$ (see (\ref{efbp})). Since the entries of the sets $\mathcal{I}_M(k)$, $\mathcal{E}_L(k)$, and $\mathcal{O}_L(k)$ are randomly selected for each data symbol, (\ref{anba1}) can be simplified to
\begin{eqnarray}\label{bbeta}
  \beta &=& \sqrt{\frac{1}{{N_\text{T}}}} \sum_{n=0}^{N_\text{T}-1} W_n e^{j \left(\frac{N_\text{T}-1}{2}-n\right)  \frac{2\pi d}{\lambda} (\cos(\theta) - \cos(\theta_\text{R}))},
\end{eqnarray}
where $W_n$ is a Bernoulli random variable and $W_n = 1$ with probability $(M+(N_\text{T}-M)/2)/N_\text{T} = \frac{N_\text{T}+M}{2N_\text{T}}$, and $W_n = -1$ with probability $\frac{N_\text{T}-M}{2N_\text{T}}$. { { For sufficiently large $N_{\text{T}}$ and $M$, such that $0<\frac{M}{N_{\text{T}}}<1$, { {the random variable $W_n$ makes $\beta$ in (\ref{bbeta}) a sum of  IID  complex random variables.}} Invoking the central limit theorem, $\beta$ converges to a complex Gaussian random variable. Thus, it can be completely characterized by its mean and variance. }}

To derive the mean of $\beta$, we first derive  $\mathbb{E} [\Re[\beta]] $ and $\mathbb{E} [\Im[\beta]]$ to obtain $\mathbb{E} [\beta]$. From (\ref{bbeta}), the expected value of $ \Re[\beta] $  can be written as  
\begin{eqnarray}\label{erbbeta}
 \nonumber \mathbb{E} [\Re[\beta]]  \hspace{-3mm} &=& \hspace{-3mm} \mathbb{E} \bigg[ \Re\bigg [ \sqrt{\frac{1}{{N_\text{T}}}} \sum_{n=0}^{N_\text{T}-1}\hspace{-1mm} W_n e^{j \left(\frac{N_\text{T}-1}{2}-n\right)  \frac{2\pi d}{\lambda} (\cos(\theta) - \cos(\theta_\text{R}))}\bigg] \bigg]\\ \nonumber
 &=&  \frac{\mathbb{E} [W_n]}{\sqrt{N_\text{T}}} \sum_{n=0}^{N_\text{T}-1} \cos\bigg(\bigg(\frac{N_\text{T}-1}{2} -n)  \times \\ \label{betam1} &&  \frac{2\pi d}{\lambda} (\cos(\theta) - \cos(\theta_\text{R})\bigg)\bigg).
\end{eqnarray}
Note that 
\begin{eqnarray}
\mathbb{E} [W_n] &=& \text{Pr}(W_n=1)-\text{Pr}(W_n=-1) \\ \label{ew} &=& \frac{N_\text{T}+M}{2N_\text{T}}-\frac{N_\text{T}-M}{2N_\text{T}} = \frac{M}{N_\text{T}}.
\end{eqnarray}
Using (\ref{ew}) and expanding (\ref{betam1}) we obtain
\begin{eqnarray}\label{eb1}
\nonumber && \mathbb{E} [\Re[\beta]] =  \frac{M}{N_\text{T}\sqrt{N_\text{T}}} \sum_{n=0}^{N_\text{T}-1} \cos\bigg(\left(\frac{N_\text{T}-1}{2} -n\right) \times \\ \nonumber && \frac{2\pi d}{\lambda} (\cos(\theta) - \cos(\theta_\text{R}))\bigg)\\ 
\nonumber &&= \frac{M}{N_\text{T}^{\frac{3}{2}}}  \bigg( \sum_{n=0}^{N_\text{T}-1} \cos\left({(N_\text{T}-1)  \frac{\pi d}{\lambda} (\cos(\theta) - \cos(\theta_\text{R}))}\right) \times \\ \nonumber && \cos\left({n \frac{2\pi d}{\lambda} (\cos(\theta) - \cos(\theta_\text{R}))}\right)  + \\ \nonumber  
&&\sum_{n=0}^{N_\text{T}-1} \sin \left({(N_\text{T}-1)  \frac{\pi d}{\lambda} (\cos(\theta) - \cos(\theta_\text{R}))}\right) \times \\  \label{eb1} &&\sin\left({n \frac{2\pi d}{\lambda} (\cos(\theta) - \cos(\theta_\text{R}))}\right) \bigg).
\end{eqnarray}
%
%
%\begin{eqnarray}\label{erbbeta2}
%\mathbb{E} [\Re[\beta]]  &=&  \frac{M}{N_\text{T}\sqrt{N_\text{T}}} \sum_{n=0}^{N_\text{T}-1} \cos\left({\left(\frac{N_\text{T}-1}{2} -n\right)  \frac{2\pi d}{\lambda} (\cos(\theta) - \cos(\theta_\text{R}))}\right)\\
%\nonumber &=& \frac{M}{N_\text{T}\sqrt{N_\text{T}}}  \bigg( \sum_{n=0}^{N_\text{T}-1} \cos\left({(N_\text{T}-1)  \frac{\pi d}{\lambda} (\cos(\theta) - \cos(\theta_\text{R}))}\right)\cos\left({n \frac{2\pi d}{\lambda} (\cos(\theta) - \cos(\theta_\text{R}))}\right) \\ \label{eb1}
%&& + \sum_{n=0}^{N_\text{T}-1} \sin \left({(N_\text{T}-1)  \frac{\pi d}{\lambda} (\cos(\theta) - \cos(\theta_\text{R}))}\right)\sin\left({n \frac{2\pi d}{\lambda} (\cos(\theta) - \cos(\theta_\text{R}))}\right) \bigg).
%\end{eqnarray}
The summation in (\ref{eb1}) can be written in closed form as
\begin{eqnarray}\label{eb30}
\nonumber \mathbb{E} [\Re[\beta]]  &=&   \frac{M  \sin(zN_\text{T}/2)}{N_\text{T}\sqrt{N_\text{T}} \sin(z/2)}  \times \\ \nonumber && \bigg(   \cos^2({z(N_\text{T}-1)/2}) +\sin^2 ({z(N_\text{T}-1)}/2)\bigg)\\ \label{eb301}
&=& \frac{M  \sin(\upsilon N_\text{T}/2)}{N_\text{T}\sqrt{N_\text{T}} \sin(\upsilon/2)},
\end{eqnarray}
where $\upsilon=\frac{2\pi d}{\lambda} (\cos(\theta) - \cos(\theta_\text{R}))$.

Similarly,  the expected value of $\Im[\beta]$ can be written as  
\begin{eqnarray}\label{eib00}
\nonumber && \hspace{-7mm} \mathbb{E} [\Im[\beta]] = \frac{M}{N_\text{T}\sqrt{N_\text{T}}} \sum_{n=0}^{N_\text{T}-1} \sin\bigg(\left(\frac{N_\text{T}-1}{2} -n\right)  \times \\ \nonumber && \frac{2\pi d}{\lambda} (\cos(\theta) - \cos(\theta_\text{R}))\bigg)\\
\nonumber &=& \frac{M}{N_\text{T}\sqrt{N_\text{T}}}  \bigg( \sum_{n=0}^{N_\text{T}-1} \sin\left({(N_\text{T}-1)  \frac{\pi d}{\lambda} (\cos(\theta) - \cos(\theta_\text{R}))}\right)  \\ \nonumber && \times \cos\left({n \frac{2\pi d}{\lambda} (\cos(\theta) - \cos(\theta_\text{R}))}\right) - \\ \nonumber
&& \sum_{n=0}^{N_\text{T}-1} \cos \left({(N_\text{T}-1)  \frac{\pi d}{\lambda} (\cos(\theta) - \cos(\theta_\text{R}))}\right) \times \\ \label{ieb10} && \sin\left({n \frac{2\pi d}{\lambda} (\cos(\theta) - \cos(\theta_\text{R}))}\right) \bigg).
\end{eqnarray}
Expanding the term (\ref{ieb10}) we obtain
\begin{eqnarray}\label{eib0}
 \hspace{-3mm}\nonumber \mathbb{E} [\Im[\beta]]   \hspace{-2mm}&=& \hspace{-2mm}   \frac{M  \sin(\upsilon N_\text{T}/2)}{N_\text{T}\sqrt{N_\text{T}} \sin(\upsilon /2)} \times \\ \nonumber && \bigg( \cos( \upsilon (N_\text{T}-1)/2 \sin( \upsilon (N_\text{T}-1)/2  ) - \\ \label{ieb1} &&  \cos( \upsilon (N_\text{T}-1)/2 \sin( \upsilon (N_\text{T}-1)/2  )   \bigg)  = 0.
\end{eqnarray}
From (\ref{eb301}) and (\ref{ieb1}) we obtain
\begin{eqnarray}
\nonumber \mathbb{E} [\beta]  &=&  \mathbb{E} [\Re[\beta]] + \mathbb{E} [\Im[\beta]] \\ \label{eb0} &=& \frac{M  \sin(N_\text{T}\frac{\pi d}{\lambda} (\cos(\theta) - \cos(\theta_\text{R})))}{N_\text{T}\sqrt{N_\text{T}} \sin(\frac{2\pi d}{\lambda} (\cos(\theta) - \cos(\theta_\text{R})))}.
\end{eqnarray}

%In this Appendix, we derive the variance of the far field radiation pattern $\beta$ in (\ref{bbeta}) assuming large number of antennas $N\text{T}$ and $\theta_\text{R} \not = \theta_\text{E}$. 

To derive the variance of $\beta$, we decompose $\beta$ into its real and imaginary parts as follows
\begin{eqnarray} 
 && \hspace{-10mm} \text{var}[\beta] \label{vea}   =   \text{var}[\Re[\beta]] + \text{var}[\Im[\beta]]\\ \nonumber
&& \hspace{-10mm} =   \text{var}\bigg[ \Re\bigg [ \sqrt{\frac{1}{{N_\text{T}}}} \sum_{n=0}^{N_\text{T}-1} W_n e^{j \left(\frac{N_\text{T}-1}{2}-n\right)  \frac{2\pi d}{\lambda} (\cos(\theta) - \cos(\theta_\text{R}))}\bigg] \bigg] +\\ && \hspace{-10mm}   \text{var}\bigg[ \Im \bigg [ \sqrt{\frac{1}{{N_\text{T}}}} \sum_{n=0}^{N_\text{T}-1} W_n e^{j \left(\frac{N_\text{T}-1}{2}-n\right)  \frac{2\pi d}{\lambda} (\cos(\theta) - \cos(\theta_\text{R}))}\bigg] \bigg] .
\end{eqnarray}
The variance of the real component can be expressed as   
 \begin{eqnarray}\label{var2}
\nonumber   \text{var}[\Re[\beta]]  &=&  \frac{1}{ N_\text{T}} \text{var} \bigg [\sum_{n=0}^{N_{\text{T}}-1} W_n \cos \bigg(\left(\frac{N_{\text{T}}-1}{2}-n\right)   \times \\ \nonumber &&  \frac{2\pi d}{\lambda}  \bigg(  \cos (\theta)-\cos (\theta_{\text{R}})   \bigg)\bigg)\bigg]
 \\ \nonumber  &=&    \frac{1}{ N_\text{T}} \sum_{n=0}^{N_{\text{T}}-1} \text{var}[W_n] \cos^2 \bigg(\left(\frac{N_{\text{T}}-1}{2}-n\right) \times \\   \label{lev} &&  \frac{2\pi d}{\lambda}  \bigg(  \cos (\theta)-\cos (\theta_{\text{R}})   \bigg)\bigg)
%&=&  \frac{\text{var}[W_n]}{N_{\text{T}}} \sum_{n=0}^{N_{\text{T}}-1} \cos^2 \bigg(\left(\frac{N_{\text{T}}-1}{2}-n\right) \frac{2\pi d}{\lambda}  \bigg( \cos \theta_{\text{R}}-\cos \theta_{\text{E}}  \bigg)\bigg),
\end{eqnarray}    % var[z] = 1-m^2/n^2
The value of var$[W_n]$ in (\ref{lev}) can be derived as 
\begin{eqnarray}
\text{var}[W_n]  &=&  \mathbb{E} [W_n^2] - (\mathbb{E} [W_n])^2\\ \nonumber
&=&   \text{Pr}(W_n=1)(1)^2+\text{Pr}(W_n=-1)(-1)^2 -  \\ \nonumber && (\mathbb{E} [W_n])^2\\
&=& \frac{N_\text{T}+M}{2N_\text{T}}+\frac{N_\text{T}-M}{2N_\text{T}} - \left(\frac{M}{N_\text{T}}\right)^2 \\ \label{vw}
&=&  \frac{N_{\text{T}}^2-M^2}{N^2_{\text{T}}}.
\end{eqnarray}    
Substituting the value of var$[W_n]$ in (\ref{lev}) and using trigonometric identities we obtain
\begin{eqnarray}
\hspace{-8mm} \nonumber \text{var}[\Re[\beta]] \hspace{-2mm}  &=&  \hspace{-2mm}  \frac{N_{\text{T}}^2-M^2}{2N^3_{\text{T}}} \times \\ \label{eva} && \bigg(N_{\text{T}}+  \frac{\sin \bigg(\frac{2\pi dN_{\text{T}}}{\lambda}  \bigg(  \cos (\theta)-\cos (\theta_{\text{R}})   \bigg)  \bigg)}{\sin \bigg(\frac{2\pi d}{\lambda}  \bigg(  \cos (\theta)-\cos (\theta_{\text{R}})   \bigg) \bigg)}   \bigg).
\end{eqnarray}
Similarly, the variance of the imaginary component of $\Im[\beta]$ can be derived as
\begin{eqnarray}
\hspace{-6mm} \nonumber  \text{var}[\Im [\beta]]  \hspace{-2mm} &=& \hspace{-2mm}   \frac{N_{\text{T}}^2-M^2}{2N^3_{\text{T}}}  \times \\ \label{eva2} && \hspace{-2mm}  \bigg(N_{\text{T}}-  \frac{\sin \bigg(\frac{2\pi dN_{\text{T}}}{\lambda}   \bigg(  \cos (\theta)-\cos (\theta_{\text{R}})   \bigg)  \bigg)}{\sin \bigg(\frac{2\pi d}{\lambda}  \bigg(  \cos (\theta)-\cos (\theta_{\text{R}})   \bigg) \bigg)}   \bigg).
\end{eqnarray}
Substituting (\ref{eva}) and (\ref{eva2}) in (\ref{vea}) yields
\begin{eqnarray}\label{var5}
 \text{var}[\beta] = \frac{N_{\text{T}}^2-M^2}{N^2_{\text{T}}}.
\end{eqnarray}

%%%%%%%%%%%%%%%%%%%%%%%%%%%%%%%%%%%%%%%%%%%%%%%%%%%%%%%%%%%%%%%%%%%%%%%%%%%%%%%%%%%%%%%%%%%%%%%%%%%%%%%%%%%%%%
%\section*{Acknowledgement} \label{sec:ackn}
%%%%%%%%%%%%%%%%%%%%%%%%%%%%%%%%%%%%%%%%%%%%%%%%%%%%%%%%%%%%%%%%%%%%%%%%%%%%%%%%%%%%%%%%%%%%%%%%%%%%%%%%%%%%%%
%%This work was sponsored by the Texas Department of Transportation under Project 0-6877 entitled “Communications and Radar-Supported Transportation Operations and Planning (CAR-STOP)”.
%
%This research was partially supported by the U.S. Department of Transportation through the Data-Supported Transportation Operations and Planning (D-STOP) Tier 1 University Transportation Center and by the Texas Department of Transportation under Project 0-6877 entitled “Communications and Radar-Supported Transportation Operations and Planning (CAR-STOP).

%\newpage
{}
\end{document}